\documentclass[letterpaper,twocolumn,10pt]{article}
\usepackage{usenix-2020-09}

\usepackage{tikz}

\usepackage{xspace}
\usepackage{balance}
\usepackage{booktabs}
\usepackage{bbm}
\usepackage{graphicx}
\usepackage{multirow}
\usepackage{float}
\usepackage{tabularx}
\usepackage{amsmath,amssymb,amsfonts}
\usepackage{xfrac} % Include xfrac package for slanted fractions
\usepackage[linesnumbered,ruled]{algorithm2e}
\usepackage{algorithmic}
\usepackage{setspace} % Include setspace package for line spacing

\newcommand\fw[1]{$\mathsf{#1}$\xspace}
\newcommand\fwbold[1]{$\boldsymbol{\mathsf{#1}}$\xspace}

\newcommand{\ceil}[1]{\left\lceil #1 \right\rceil}

\newcommand{\mtrx}[1]{\uppercase{\mathbf{#1}}}
\newcommand{\vctr}[1]{\lowercase{\mathbf{#1}}}

\newcommand{\OT}[2]{\binom{#1}{1}\text{--}\mathsf{OT}_{#2}}

\newcommand{\ROT}[2]{\binom{#1}{1}\text{--}\mathsf{ROT}_{#2}}

\newcommand{\COT}[2]{\mathsf{COT}_{#2}}
\newcommand{\COTsnd}[2]{\mathsf{COT}_{#2}\text{--}\mathsf{send}}
\newcommand{\COTrec}[2]{\mathsf{COT}_{#2}\text{--}\mathsf{receive}}

\newcommand{\sshr}[3]{\langle #1 \rangle_{#2}^{#3}}
\newcommand{\prtcl}[1]{\mathcal{F}_{\mathsf{#1}}}

\newcommand{\poly}[1]{\overline{\lowercase{\mathbf{#1}}}}
\newcommand{\pt}[1]{\poly{pt}_\mathbf{#1}}
\newcommand{\ct}[1]{\overline{\mathbf{ct}}_\mathbf{#1}}
\newcommand{\ek}[1]{\overline{\mathbf{ek}}_\textrm{#1}}

\SetKwInput{KwInput}{Input}                % Set the Input
\SetKwInput{KwOutput}{Output}              % set the Output

\begin{document}
%-------------------------------------------------------------------------------
%-------------------------------------------------------------------------------
%don't want date printed
\date{}

\title{\fwbold{SecONNds}: Secure Outsourced Neural Network Inference on ImageNet}

% for single author (just remove % characters)
\author{
{\rm Shashank Balla}\\
Independent\\
shashank.raghavaballa@gmail.com
% \and
% {\rm Second Name}\\
% Second Institution
% copy the following lines to add more authors
% \and
% {\rm Name}\\
%Name Institution
} % end author
%-------------------------------------------------------------------------------
%-------------------------------------------------------------------------------
\maketitle
%-------------------------------------------------------------------------------
%-------------------------------------------------------------------------------
\begin{abstract}
%-------------------------------------------------------------------------------
%-------------------------------------------------------------------------------
The widespread adoption of outsourced neural network inference presents significant privacy challenges, as sensitive user data is processed on untrusted remote servers. Secure inference offers a privacy-preserving solution, but existing frameworks suffer from high computational overhead and communication costs, rendering them impractical for real-world deployment. We introduce \fw{SecONNds}, a non-intrusive secure inference framework optimized for large ImageNet-scale Convolutional Neural Networks. \fw{SecONNds} integrates a novel fully Boolean Goldreich-Micali-Wigderson (GMW) protocol for secure comparison -- addressing Yao's millionaires' problem -- using preprocessed Beaver's bit triples generated from Silent Random Oblivious Transfer. Our novel protocol achieves an online speedup of 17$\times$ in nonlinear operations compared to state-of-the-art solutions while reducing communication overhead. To further enhance performance, \fw{SecONNds} employs Number Theoretic Transform (NTT) preprocessing and leverages GPU acceleration for homomorphic encryption operations, resulting in speedups of 1.6$\times$ on CPU and 2.2$\times$ on GPU for linear operations. We also present \fw{SecONNds\text{-}P}, a bit-exact variant that ensures verifiable full-precision results in secure computation, matching the results of plaintext computations. Evaluated on a 37-bit quantized SqueezeNet model, \fw{SecONNds} achieves an end-to-end inference time of 2.8 s on GPU and 3.6 s on CPU, with a total communication of just 420 MiB. \fw{SecONNds}' efficiency and reduced computational load make it well-suited for deploying privacy-sensitive applications in resource-constrained environments. \fw{SecONNds} is open source and can be accessed from: \url{https://github.com/shashankballa/SecONNds}.
\end{abstract}
%-------------------------------------------------------------------------------
%-------------------------------------------------------------------------------
\section{Introduction}
%-------------------------------------------------------------------------------
%-------------------------------------------------------------------------------
Machine learning (ML) has become ubiquitous, with pre-trained neural networks (NN) playing a pivotal role in numerous applications that shape our daily interactions, such as image recognition, natural language processing, and recommendation systems. To handle the computational demands of these models, especially large ones such as Deep Neural Networks (DNN), it is common practice to outsource computations to remote cloud servers. This approach allows resource-constrained devices, such as mobile and embedded systems, to leverage powerful models by offloading the heavy computation and receiving the final result. However, this practice presents significant privacy risks, as sensitive user data is processed on remote servers that may not be fully trusted.

\begin{figure}[t]
    \centering
    \includegraphics[width=\columnwidth]{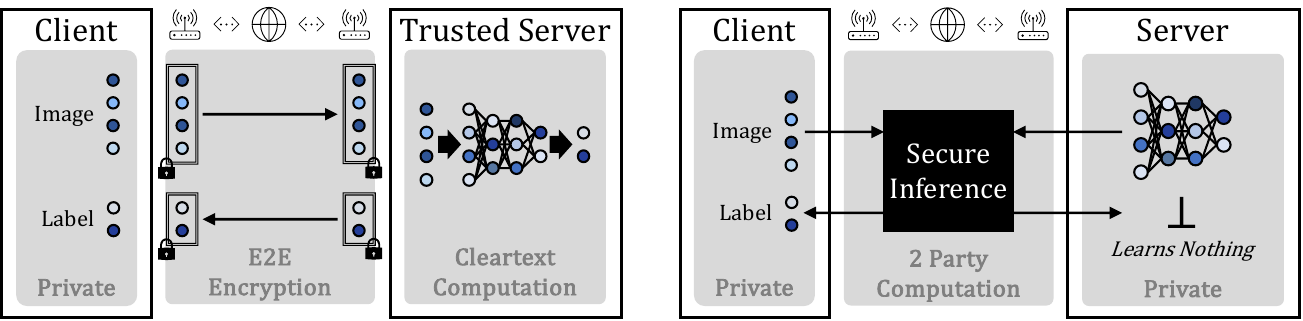}
    \vspace{5mm}
    \caption{Orthodox Outsourced Inference (left) vs. Secure Inference (right)}
    \label{fig:secinf}
\end{figure}

In response to these privacy concerns, the security and privacy research community has introduced frameworks for \emph{Secure Inference}, illustrated in Figure~\ref{fig:secinf}. Using cryptographic techniques, these frameworks ensure the protection of all private data from all parties involved: users' inputs and outputs of the NN inference, and the proprietary model parameters of the service provider. These solutions allow users to infer on pre-trained NNs without ever exposing their data.

Neural networks are diverse in architecture, but are essentially composed of an alternating series of multidimensional \emph{linear} operations and \emph{nonlinear} operations which are mostly unary. Convolutional Neural Networks (CNN) are one of the most popular classes of architectures that have become essential for computer vision tasks. CNNs feature multidimensional convolution operations to match learned spatial features with the input image, ReLU operations to pick the task-related matches, and pooling operations to downsample the matches. CNNs started the deep learning era ~\cite{krizhevsky12AlexNet} with models growing larger and larger in terms of parameters.

For efficient and practical secure inference, it is crucial to design computation algorithms for low-level neural network operations with efficient cryptographic primitives that provide privacy, considering the resource allocation of both parties. Employing a zero-trust model for secure inference requires that the framework does not reveal any private data and involves only the user and the service provider, necessitating a purely $2$-party security model. This necessitates cryptographic primitives for \emph{Secure 2-Party Computation} (2PC) protocols: Oblivious Transfer (OT)\cite{rabin1981ot}, Garbled Circuit (GC)\cite{yao1982protocols}, Goldreich-Micali-Wigderson (GMW)\cite{goldreich1987play}, and Homomorphic Encryption (HE)~\cite{gentry09fhe}.

OT allows for secure exchange of messages and enables secure Look-Up Table (LUT) evaluations. Similarly to how FPGA boards are programmed, OT is universal in its expressiveness and serves as a foundation for GC and GMW. GC is a one-round protocol for securely evaluating boolean circuits and was the first solution proposed for 2PC. GMW enables secure evaluation of arithmetic circuits on fixed-point data on top of boolean circuits, but it is a highly interactive protocol that requires constant communication between parties for every operation. Both GC and GMW assume symmetric resource allocation for both parties and necessitate an equal amount of local computation on both ends.

HE is an encryption scheme that preserves some algebraic structure in the encrypted data, allowing computations on ciphertexts that reflect as simple operations on the underlying plaintexts. While partial HE schemes~\cite{rivest78data, paillier99public} offer a limited set of operations, fully HE schemes~\cite{gentry09fhe, ducas14fhew, chillotti18tfhe} enable arbitrary computations but are very slow. 

Leveled HE schemes~\cite{bgv11, brakerski11bfv, fan12bfv, cheon17ckks} strike a good balance by supporting arithmetic circuits ($+$, $\times$) on fixed-point data up to a fixed size (multiplicative depth) with usable performance. The main advantage of HE is in securely outsourcing computation; the user sends encryptions to the server, which performs all computations and returns the encrypted result to the user. This makes HE best suited to asymmetric scenarios involving users with resource-constrained devices and a service provider with a powerful cloud server.

2PC primitives (GC, GMW, HE) are tailored for fixed-point circuits, so it is common to convert neural network operations to fixed-point representations. This task has been thoroughly researched in machine learning through Quantized Neural Networks (QNN)~\cite{hubara16qnn}. QNNs use fixed-point representations for all data, including the trained model parameters. 

Almost all of the trained parameters of a model belong to its linear layers, and are only used for computing linear combinations of the inputs. Therefore, the size of a model directly corresponds to the number of linear operations involved. With GC or GMW, the communication footprint of linear layers scales with the number of total scalar multiplications. In contrast, with HE, the communication footprint scales only with the sizes of the input and output vectors. Today, leveled HE outperforms any other 2PC primitive in evaluating high-dimensional linear algebra operations on encrypted data~\cite{balla23heliks}. 

Nonlinear operations make up the remainder of the computation and are indispensable in neural networks. These nonlinear operations involve comparisons, and in fact, this was one of the first problems studied in secure computation by Andrew Yao~\cite{yao1982protocols}, who dubbed it the Millionaires’ Problem, $\mathsf{MILL} = \mathbbm{1}\{i_0 > i_1\}$ which returns $1$ if party-$0$’s input ($i_0$) is greater than party-$1$‘s input ($i_1$). The bulk of online runtime of these operations in state-of-the-art secure inference protocols comes from the secure comparisons (Millionaires’ protocol). 

Solving the Millionaires’ Problem involves securely evaluating boolean operations which requires GC or GMW. GC involves a ciphertext expansion of $O(\lambda b)$\footnote{$\lambda$ is the computational security parameter, typically $128$} for comparing a pair of $b$ bit secrets. A GMW protocol initialized with Silent OT Extension~\cite{boyle19SilentOT} performs the same task with an overhead of only $O(b)$, greatly propelling its performance over GC. However, existing GMW protocols still incur high computational overhead, especially for large-scale datasets like ImageNet, making them impractical for real-world applications.

\textbf{Our Approach:} We present \fwbold{SecONNds}, a non-intrusive secure inference framework that develops new algorithms for the Millionaires’ Problem within the GMW protocol and enhances HE-based linear operations using Number Theoretic Transform preprocessing and GPU acceleration. By holistically improving both nonlinear comparisons and linear computations, \fw{SecONNds} significantly boosts the performance of secure inference, making it practical for large-scale applications like ImageNet. Our main contributions are as follows:

\noindent
\textbf{+ New solutions to the Millionaires’ Problem:} We present $\prtcl{MILL}$ for secure comparisons, a fully Boolean GMW protocol with Beaver’s bit triples (triples)\cite{beaver1996correlated} generated using silent Random OT (ROT)\cite{kang20ferret} that achieves faster runtimes and lower communication than prior work, and an alternate variant that incurs only a logarithmic number of rounds, with slightly higher computation and communication costs. Both are further enhanced with an offline triple buffer and a chunked generator optimized for silent OT.

\noindent
\textbf{+ Efficient protocols for neural network operations:} We develop new 2PC protocols for ReLU, Max Pooling, and Truncation using $\prtcl{MILL}$ with offline triples. For linear algebra operations, we employ the BFV HE scheme~\cite{brakerski11bfv, fan12bfv} featuring one-time preprocessing with Number Theoretic Transform (NTT)~\cite{balla23heliks} and server-side GPU acceleration~\cite{troy}. Our protocols achieve online speedups of $17\times$ for Max Pooling, $11\times$ for ReLU, $2.1\times$ for Truncation, and $2.2\times$ for convolution with HE on GPU, over prior art.

\noindent
\textbf{+ End-to-end implementation and evaluation:} We evaluate our open-source implementation on a $37$ bits quantized SqueezeNet model~\cite{iandola16squeezenet} for ImageNet~\cite{deng09imagenet}. \fw{SecONNds} achieves an E2E runtime of just $2.8$ seconds on GPU and $3.6$ seconds on CPU for secure inference, involving $420$ MiB of total communication. Its E2E performance is $4.2\times$ faster online compared to the state-of-the-art. We also implement \fw{SecONNds\text{-}P}, a bit-exact variant that ensures full-precision, verifiable results in secure computation compared to plaintext.
%-------------------------------------------------------------------------------
%-------------------------------------------------------------------------------
\section{Background}
\label{background}
%-------------------------------------------------------------------------------
%-------------------------------------------------------------------------------
\subsection{Mathematical Notation}
\label{sec:math}
%-------------------------------------------------------------------------------
In this section, we introduce the mathematical notation used throughout this paper. Integer vector fields are denoted by $\mathbb{Z}^N_Q$ where $N$ is the dimensionality of the vector space and $Q$ is the field modulus. Polynomial rings are denoted by $\mathcal{R}^N_{\,\,\,Q} = \mathbb{Z}_Q[X] \bmod \big(X^N+1\big)$ for polynomials of degree less than $N$ with coefficients from $\mathbb{Z}_Q$. Here $N$ is the polynomial degree modulus and $Q$ is the coefficient modulus.

Scalars in $\mathbb{Z}_N$ are denoted by normal text $x$ or $Y$. Vectors in $\mathbb{Z}^m_N$ are denoted by bold lowercase letters $\vctr{x}$, and matrices in $\mathbb{Z}^{p\times q}_N$ by bold uppercase letters $\mtrx{X}$. A polynomial in $\mathcal{R}^N_{\,\,\,Q}$ is denoted by a bold lowercase letter with an overline $\poly{p}$. A HE plaintext encoding a secret $\vctr{m}$ is denoted as $\pt{m}$ and its ciphertext is denoted as $\ct{m}$.

Bitwise complement of a scalar $x$ is denoted by $x'$. Operators $\oplus$ and $\land$ are reserved for addition (XOR) and multiplication (AND) in $\mathbb{Z}_2$. For $p\!\!\in\!\!\{0$:\,Server$, 1$:\,Client$\}$, party $\mathcal{P}_p$'s linear secret share of $i$ over $\mathbb{Z}_N$ is denoted by $\sshr{i}{p}{N}$, i.e. $i = \sshr{i}{0}{N} + \sshr{i}{1}{N} \bmod N$. The indicator function, returning $1$ if \emph{condition} is satisfied or else $0$, is denoted by $\mathbbm{1}\{ \text{\emph{condition}} \}$.
%-------------------------------------------------------------------------------
\subsection{Secure 2-Party Computation (2PC)} 
\label{sec:seccomp}
%-------------------------------------------------------------------------------
Secure 2-party computation enables two parties to jointly compute a function over their private inputs while keeping them confidential. Different cryptographic primitives support 2PC, each offering unique trade-offs in terms of efficiency, communication overhead, and computational complexity.

\textbf{Oblivious Transfer (OT)}~\cite{rabin1981ot} is a cryptographic primitive that allows a sender to transfer one out of many pieces of information to a receiver without knowing which piece was transferred, and ensuring nothing is learned about the other pieces. There are various types of OT based on functionality.

\emph{Correlated OT (COT)}: In a Correlated OT, denoted by $\COT{2}{b}$, the sender inputs a $b$ bit string $m$, and the receiver obtains a $b$ bit string $m_c = m + c \cdot \delta$, where $\delta$ is a fixed $b$ bit correlation known to the sender, and $c \in \mathbb{Z}_2$ is the receiver's choice bit. The outputs are correlated according to the sender's input $\delta$. COT is a foundational protocol, commonly used for generating correlated randomness. 

\emph{Random OT (ROT)}: In a Random OT, the sender's messages are randomly generated, and the receiver obtains one of them based on its choice bit. In a $\ROT{N}{b}$, the sender obtains $N$ random $b$ bit strings $\{r_0, ..., r_{N-1}\}$, and the receiver obtains one string $r_c$, where $c \in \mathbb{Z}_N$ is the receiver's choice bit. ROT is often used for generating symmetric cryptographic keys or masks in secure protocols.

\emph{Chosen OT}: This is the most expressive form of OT and is often simply referred to as OT. In a $\OT{N}{b}$, the sender inputs $N$ messages $\{m_0, ..., m_{N-1}\}$, and the receiver only learns $m_c$ where $c \in \mathbb{Z}_N$ is its choice. Chosen OT is essential when one party needs to send specific messages to the other based on the receiver's choice. For example, to evaluate a secure LUT, the sender sets the values of the LUT, the receiver indexes with its private choice and only learns the output value.

These OT primitives can be implemented in various styles based on the requirements. Efficient OT extension protocols~\cite{ishai2003extending, boyle19SilentOT} offer the best communication performance. They allow a large number of OTs to be generated from a small number of base OTs, significantly improving efficiency. We include a detailed discussion on OT extension in Appendix~\ref{sec:ot}.

\textbf{Garbled Circuits (GC)}~\cite{yao1982protocols} is a 2PC protocol where one party (the garbler) encrypts a Boolean circuit, allowing the other party (the evaluator) to compute it without learning intermediate values or inputs. Each wire is assigned two $\lambda$-bit random keys (labels) for logical 0 and 1. The garbler encrypts each gate's truth table and sends it, along with the wire labels of its own inputs, to the evaluator. The evaluator obtains the wire labels for its inputs using $\COT{2}{\lambda}$, ensuring the garbler remains unaware of the evaluator's inputs. It then uses these labels to sequentially decrypt each gate, ultimately obtaining the output wire labels. While GC provides security with a single communication round, it can incur high computational and communication costs due to the need to encrypt every gate in the circuit and handle large garbled tables where each bit is represented with a large $\lambda$ bit ciphertext.  

\textbf{Goldreich-Micali-Wigderson (GMW)}~\cite{goldreich1987play} is an interactive protocol that can be built with Linear Secret Sharing Schemes (LSSS)~\cite{cramer00lsss}. Each party $\mathcal{P}_p$ holds a secret share $\sshr{i}{p}{N}$ for every private input $i$ in the circuit. For secure additions (XOR gates), computation is performed locally without interaction, while secure multiplications (AND gates) require secure communication between the parties, by means of OT, to obtain secret shares of the product. The communication volume of GMW scales with the number of multiplication operations, requiring one round per multiplication. However, techniques such as circuit randomization and the use of correlated pseudorandomness, specifically through Beaver's triples, can significantly reduce the round complexity. 

\textbf{Beaver's Bit Triples}~\cite{beaver92efficient, beaver1996correlated} enable fast computation of an AND gate without relying on Chosen OT during the online phase. A Beaver's bit triple consists of $\big\{\sshr{a}{p}{2}, \sshr{b}{p}{2}, \sshr{c}{p}{2}\big\}$, where $a$ and $b$ are random bits, and $c = a \land b$. This can be generated offline using 2 calls to $\ROT{2}{1}$. During the online phase, given secret-shared inputs $\sshr{x}{p}{2}$ and $\sshr{y}{p}{2}$, the parties compute local corrections with the pre-shared triple values and exchange these correction bits to adjust their output shares (illustrated in Algorithm~\ref{algo:and} of Appendix~\ref{sec:secand}).

While both GC and GMW offer the ability to perform any computation, the primary challenge with them is the nearly equal computational effort demanded from both parties, and hence they do not offer the ability to outsource computation. 

\textbf{Homomorphic Encryption (HE)} allows clients to encrypt data locally, outsource computation to a server that operates on encrypted data, and then decrypt the results themselves. This reduces communication overhead significantly, as only input and output data are transferred, while the server handles most of the computation. Fully Homomorphic Encryption (FHE), such as Gentry's scheme \cite{gentry09fhe}, TFHE \cite{chillotti18tfhe}, or FHEW \cite{ducas14fhew}, offers unlimited encrypted computation through bootstrapping but incurs large runtimes. Leveled HE uses large input ciphertexts and foregoes bootstrapping, which limits encrypted computation and significantly improves runtimes. BGV \cite{bgv11}, BFV \cite{brakerski11bfv, fan12bfv}, and CKKS \cite{cheon17ckks} are popular Leveled HE schemes based on the \emph{Ring Learning With Errors} (RLWE)~\cite{lyubashevsky13rlwe} hard problem.

RLWE HE schemes support addition, multiplication, and rotation operations on ciphertexts. HE multiplications have large noise growth, affecting the amount of encrypted computation, while additions and rotations have a negligible effect. In terms of runtimes, rotations are the slowest due to a key-switching procedure with special public evaluation keys $\ek{}$, followed by multiplications which involve $O(N^2)$ convolution operations, while additions are relatively fast. HE multiplications can be optimized using NTT, which reduces its complexity to $O(N\log{N})$. We discuss more about RLWE HE and the role of NTT in Appendix~\ref{sec:appendix_he}.

\begin{table}[t]
    \caption{2PC Performance for $N \times N$ Matrix-Vector Product}
    \label{table:seccomp}
    \centering
    \resizebox{\columnwidth}{!}{%
        \begin{tabular}{ccccccc}
            \toprule
            \textbf{2PC} & \multicolumn{3}{c}{\textbf{Offline}} & \multicolumn{3}{c}{\textbf{Online}} \\ 
            % \cline{2-7} 
            \textbf{Protocol}& Rounds  & Comm.  & Comp.   & Rounds  & Comm.  & Comp.  \\ 
            \midrule
            GC  & $0$   & $0$   & Server & $1$   & $O(\lambda N^2)$ & Client \\
            GMW & $1$ & $O(N^2)$ & Both  & $1$ & $O(N^2)$ & Both   \\
            HE  & $0$     & $0$     & Server & $1$   & $O(N)$   & Server \\
            \bottomrule
        \end{tabular}
    }
\end{table}

\textbf{2PC Cost Comparison.} Table~\ref{table:seccomp} highlights the costs of different 2PC protocols -- GC, GMW with Beaver's triples, and HE -- for performing a secure matrix-vector product of size $N \times N$. In this setting, the client holds a private input vector, and the server holds a private input matrix (e.g., trained weights in a Support Vector Machine model). The goal is for the client to obtain the output vector without revealing its input or learning the server's matrix.

GC requires the server to garble the circuit locally in the offline phase, incurring computational costs. This offline phase doesn't require any communication. In the online phase, the server sends the garbled circuit to the client, resulting in a communication cost proportional to $\lambda N^2$, where $\lambda$ is the security parameter. Additionally, $N$ Correlated OTs are used to share the input wire labels corresponding to the client's input vector during the online phase. The client then evaluates the garbled circuit using its input labels and learns the output.

GMW involves generating $N^2$ Beaver's triples using $2N^2$ Random OTs during the offline phase, $1$ triple per scalar multiplication required in the matrix-vector product. Both parties incur symmetric computational and communication costs during this phase. In the online phase, the parties exchange correction bits in a single round of communication, leading to a communication cost proportional to $N^2$ and maintaining symmetric computation between both parties.

HE, during the offline phase, may involve the server preprocessing its matrix with NTT~\cite{balla23heliks}. In the online phase, the client encrypts its input vector and sends it to the server. The server performs the matrix-vector multiplication with HE using its private matrix and returns the encrypted output vector to the client. The communication cost is proportional to $N$, and the computational burden is primarily on the server, which performs the homomorphic operations locally.
%-------------------------------------------------------------------------------
\subsection{Secure Inference Frameworks} 
\label{sec:secinf}
%-------------------------------------------------------------------------------
Many major tech companies, such as Apple~\cite{apple_private_cloud_compute}, Google~\cite{google_confidential_computing}, Microsoft~\cite{azure_confidential_computing}, and Amazon~\cite{aws_nitro_enclaves}, are making a huge push into privacy-focused solutions for AI inference, but they choose to use Trusted Execution Environments (TEE). For example, Apple's Private Cloud Compute (PCC) uses a TEE built into custom Apple silicon to protect user data during cloud-based AI processing. PCC ensures data is processed securely and in a stateless manner, reducing risks associated with data retention. The main issue with such outsourced inference systems (depicted in the left half of Figure~\ref{fig:secinf}) is they protect the data with encryption only during transit; the computation is still performed in cleartext, albeit inaccessible due to hardware privileges. Despite their secure design, limitations inherent in TEEs, like potential side-channel vulnerabilities, hamper their applicability to privacy critical applications~\cite{kou23teeattack, zhang24teeattack, javeed23teeattack}. 

2PC based cryptographic methods for secure inference (depicted in the right half of Figure~\ref{fig:secinf}) do not have any dependencies and leverage strong cryptographic guarantees to protect data in all phases (locations): rest (memory), transit (network) and computation (processor). These frameworks enable neural network computations on private data, without revealing any sensitive information to any participant. Over the years, several frameworks have been developed, leveraging various cryptographic techniques to balance efficiency and privacy. 

\fwbold{CryptoNets}~\cite{gilad16cryptonets} is the first secure inference framework to run a Convolutional Neural Network on MNIST~\cite{deng12mnist}. It is constructed entirely with HE and performs secure inference in $1$ round. It uses arithmetized CNNs that are fine-tuned with matrix multiplications and polynomial activations in place of convolutions and ReLU, respectively. 

\fwbold{MiniONN}~\cite{liu17minionn} is the first non-intrusive\footnote{Does not require any model customization or fine-tuning.}, mixed-protocol framework to perform inference on CIFAR-10~\cite{krizhevsky2009cifar10}. It uses GC, GMW, and HE to implement protocols for most of the popular CNN operations. It implements exact protocols for (piecewise) linear functions like (ReLU) convolution, and smooth functions are approximated with splines (piecewise polynomials). \fw{MiniONN} features an offline preprocessing phase to set up Beaver's bit triples~\cite{beaver1996correlated} using HE. During the online phase, which involves evaluating the inference result of the private image, they use GMW with preprocessed triples for all linear operations and GC for comparisons. \fw{MiniONN}'s novel mixed protocol design requires communication after every layer, introducing additional communication rounds and yet offers two orders of magnitude better performance than \fw{CryptoNets}. All later frameworks, strictly targeting E2E performance, adhere to this mixed-protocol design.

\fwbold{Gazelle}~\cite{juvekar18gazelle} implements the first purpose-built algorithm for performing convolutions on $3$-D data using HE. It combines these HE protocols for linear layers with a GC protocol for secure comparisons in $\mathbb{Z}_P$ where $P$ is a HE plaintext prime modulus and shows an order of magnitude better performance than \fw{MiniONN} on CIFAR-10. This motivated a shift back to HE for linear layers moving forward. 

\fwbold{CrypTFlow2}~\cite{rathee20cryptflow2} introduces a new protocol for the millionaires' problem leveraging the state-of-the-art optimizations and techniques for IKNP-style OT extension~\cite{kolesnikov13kkot, asharov13more}. It features highly efficient OT-based implementations of all nonlinear operations over $\mathbb{Z}_P$ as well as $\mathbb{Z}_{2^b}$ using this millionaires' protocol. The authors observe that protocols for $\mathbb{Z}_P$ are more expensive than the corresponding protocols for $\mathbb{Z}_{2^b}$, but in order to take advantage of the gains offered by HE, they implement and integrate protocols for nonlinear operations over $\mathbb{Z}_P$ with \fw{Gazelle}'s HE protocols. \fw{CrypTFlow2} outperforms GC protocols for nonlinear operations in $\mathbb{Z}_P$ and $\mathbb{Z}_{2^b}$ by an order of magnitude, and is the first to show inference on ImageNet~\cite{deng09imagenet}. It employs \emph{faithful truncation} which ensures bit-exact results in secure computation compared to plaintext.

\fwbold{Cheetah}~\cite{huang22cheetah} implements brand new algorithms for linear algebra operations over $\mathbb{Z}_{2^b}$ using HE. It builds on the observation that HE multiplications, basically polynomial multiplications, implicitly compute vector dot products over the coefficients. To capitalize on this, \fw{Cheetah} implements an encoding scheme that bypasses the traditional encoding space of schemes like BFV/BGV and instead encodes messages directly into the coefficients of the polynomials. The messages are placed strategically within the polynomials such that one vector dot product is computed with a single polynomial multiplication. This strategy results in a sparse output with very few coefficients containing the actual result. \fw{Cheetah} addresses this by extracting relevant coefficients from the output.  \fw{Cheetah}'s protocols for linear algebra do not involve any HE rotations (slowest HE operation) and outperform \fw{Gazelle}'s protocols by $5\times$ in computation and communication. 

\fw{Cheetah} uses \fw{CrypTFlow2}'s algorithms with silent OT primitives derived from Ferret Silent OT~\cite{kang20ferret} for its nonlinear operations over $\mathbb{Z}_{2^b}$. To alleviate the computation overhead of silent OT, \fw{Cheetah} implements a $1$ bit approximate truncation particularly optimized for scenarios where the secret value is known to be positive. Truncation is delayed until after ReLU instead of performing it right after convolution or matrix multiplication to take advantage of the new optimized protocol. \fw{Cheetah} achieves $3\times$ faster E2E runtimes with less than $10\times$ communication over \fw{CrypTFlow2}.

\fwbold{HELiKs}~\cite{balla23heliks} is the latest in the line of works targeting secure linear algebra operations for CNNs on high dimensional data leveraging HE. It presents new protocols for secure linear algebra operations HE over $\mathbb{Z}_P$ that outperform corresponding protocols in \fw{Cheetah} in terms of both computation and communication for the same precision, while strictly adhering to the definitions of the HE schemes used. \fw{Cheetah}'s HE protocols for linear operations generate very sparse HE results, the coefficient extraction process deviates from HE scheme definitions and the final outputs generated by the protocols bear no similarity in structure compared to their corresponding inputs. \fw{Cheetah}'s HE results cannot be readily used by the server for any subsequent HE operations (if need be). 

\fw{HELiKs} provides modular kernels that maintain the same encoding format of the input in the output. The performance of these kernels is significantly improved by taking advantage of many HE and algorithmic optimizations, such as noise growth reduction, $1$-step rotations, NTT preprocessing, tiling for large inputs, and symmetric key encryption. \fw{HELiKs} uses the secure computation protocols of \fw{CrypTFlow2} for nonlinear operations to perform secure inference.

\textbf{Transformer Inference} with 2PC has seen a huge push in research recently with frameworks like, \fw{SIRNN}~\cite{rathee21sirnn}, \fw{Iron}~\cite{hao22iron}, \fw{BOLT}~\cite{pang23bolt}, and \fw{BumbleBee}~\cite{lu23bumblebee}. All of these transformer works reuse modules from the aforementioned frameworks as foundation protocols and use them to build higher-level operations, e.g. softmax, GeLU, attention, etc.
%-------------------------------------------------------------------------------
\subsection{Computational Bottlenecks}
\label{compbott}
\begin{figure}[t]
    \centering
    \includegraphics[width=0.95\columnwidth]{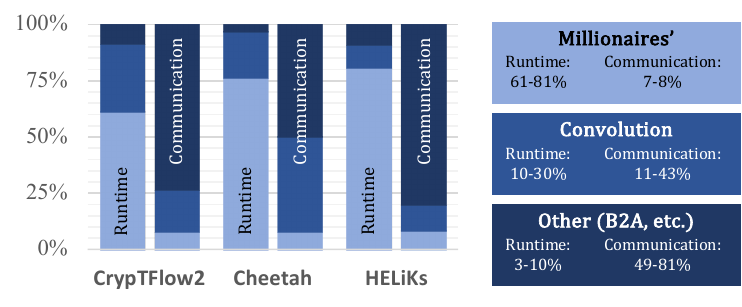}
    \caption{Secure Inference costs per protocol.}
    \label{fig:millgraph}
\end{figure}

%-------------------------------------------------------------------------------
Despite advancements, secure inference frameworks face significant computational challenges, particularly in non-linear operations involving secure comparisons. The Millionaires' Protocol, which determines if one private value is greater than another, is essential for activation functions like ReLU and operations like Max Pooling. To quantify the overhead, we evaluated the single-threaded runtime and communication costs of \fw{CrypTFlow2}, \fw{Cheetah}, and \fw{HELiKs} per protocol for running a $37$ bit SqueezeNet~\cite{iandola16squeezenet} model under the same compute and network setup (described in Section~\ref{sec:eval}). We used the same Silent OT Extension~\cite{kang20ferret} back-end for all frameworks. The graph illustrated in Figure~\ref{fig:millgraph} shows the distribution of total runtime and communication across each protocol: Millionaires' (secure comparisons), Convolution, and Other.

\textbf{Millionaires' Protocol} is the most significant contributor to runtime across all frameworks, accounting for $61-76\%$ of the total. \fw{CrypTFlow2} and \fw{HELiKs} with faithful truncation spend $140$ seconds for secure comparisons, while \fw{Cheetah} reduces this to around $60$ seconds by employing optimizations like $1$ bit approximate truncation and delayed computation. However, even with these improvements, secure comparisons remain the primary bottleneck due to their reliance on bit-level operations and multiple rounds of interaction. In terms of communication, the cost from comparisons is very low $7-8\%$, with \fw{Cheetah} requiring less than half of \fw{CrypTFlow2}.

\textbf{Convolution Operations} are the next major bottleneck, with runtime contributions ranging between $17\%$ and $30\%$. \fw{Cheetah} brings the convolution time down to $17$ seconds, compared to $70$ seconds for \fw{CrypTFlow2} and $34$ seconds for \fw{HELiKs}. Its performance improves significantly with the elimination of HE rotations and could further benefit from the optimizations introduced by \fw{HELiKs} like Number Theoretic Transform (NTT) pre-computation. Convolution operations also contribute to $11-43\%$ of the total communication. \fw{HELiKs}, with a communication cost of $124$ MiB, outperforms both \fw{CrypTFlow2} ($217$ MiB) and \fw{Cheetah} ($205$ MiB), owing to effective noise management during computation that results in smaller ciphertext sizes.

\textbf{Other Operations}, including local plaintext operations and binary-to-arithmetic share conversions (B2A), contribute modestly to the runtime across all frameworks but impose significant communication overhead. The B2A operations have large communication costs since they involve translating $1$ bit inputs to $b$ bit field elements.

\textbf{Implications for Transformers.} The computational challenges posed by secure comparisons extend beyond basic CNN operations to more complex architectures like transformers. Recent transformer secure inference frameworks heavily rely on the millionaires' protocol for range checks in activation functions. Frameworks like \fw{SIRNN}~\cite{rathee21sirnn} and \fw{BOLT}~\cite{pang23bolt} directly employ \fw{CrypTFlow2}'s protocol with IKNP-style OT extensions, while \fw{Iron}~\cite{hao22iron} and \fw{BumbleBee}~\cite{lu23bumblebee} use \fw{Cheetah}'s variant with silent-OT extension. While these frameworks introduce novel methods for approximating non-linear operations with splines -- primarily focusing on reducing the number of polynomial segments and consequently the calls to millionaires' protocol -- they maintain the fundamental comparison protocol unchanged. This widespread reliance on existing comparison protocols, coupled with their significant performance overhead shown in our analysis, highlights a critical need for new approaches to secure comparisons in privacy-preserving neural network inference.
%-------------------------------------------------------------------------------
%-------------------------------------------------------------------------------
\section{Threat Model}
\label{sec:threat}
%-------------------------------------------------------------------------------
%-------------------------------------------------------------------------------
\fw{SecONNds} operates under a two-party semi-honest (honest-but-curious) security model involving a client with private input data and a server with a private model. The framework assumes both parties execute protocols correctly but may record all observed values, no collusion between parties exists, and network adversaries can observe but not modify communications. Under this security model, the framework ensures: (1) \emph{Client Privacy}: Server learns nothing about the client's input image or inference result, (2) \emph{Model Privacy}: Client learns nothing about the model parameters beyond what can be inferred from the output label, and (3) \emph{Computational Security}: $128$ bit security parameter based on standard cryptographic assumptions. The model excludes active adversaries who may deviate from the protocol.
%-------------------------------------------------------------------------------
%-------------------------------------------------------------------------------
\section{Millionaires’ Protocol}
\label{sec:mill}
%-------------------------------------------------------------------------------
%-------------------------------------------------------------------------------
\begin{figure}[t]
    \centering
    \includegraphics[width=\columnwidth]{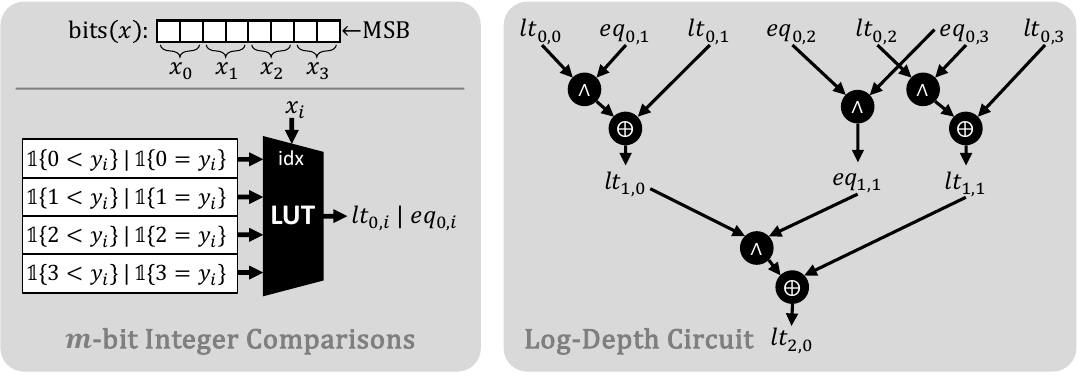}
    \caption{Millionaires' protocol using LUT (Chosen OT).}
    \label{fig:cf2mill}
\end{figure}

The Millionaires' problem was conceptualized by Yao~\cite{yao1982protocols} as two millionaires who want to learn who is richer without disclosing their wealth to each other. The solution for the millionaires' problem corresponds to securely evaluating a comparison operation involving two parties with one private input each. The current state-of-the-art algorithm for millionaires' with OT based secure evaluation was presented by Rathee et al. in \fw{CrypTFlow2}~\cite{rathee20cryptflow2}. The core part of this algorithm comes from the work of Garay et al.~\cite{garay07practical}, who proposed a novel approach of decomposing the two large $b$ bit inputs, $x$ of one party ($\mathcal{P}_0$) and $y$ of the other party ($\mathcal{P}_1$), into $q=b/m$ consecutive $m$ bit segments each, $\{x_0, ..., x_{q-1}\}, \{y_0, ..., y_{q-1}\}$, and computing the result with an arithmetic circuit over $m$ bit numbers. This algorithm is illustrated in Figure~\ref{fig:cf2mill} for $8$ bit inputs ($b=8$) with a segment size of $m=2$ bits. 

The computation begins with evaluating the inequality $lt_{0, i}\!\!= \mathbbm{1}\!\{\!x_i\!\!<\!\!y_i\!\}$, and equality $eq_{0, i}\!\! = \mathbbm{1}\!\{\!x_i\!=\!y_i\!\}$ results for all segment-pairs $i \in [0, q)$. This is followed by an arithmetic circuit that combines these results in a binary tree fashion with depth $\ceil{\log q}-1$. The results of the $q$ inequality and equality comparisons are laid out at the root level ($j=0$), and for every level, $j \geq 1$, two bits, $lt_{j, i}$ and $eq_{j, i}$, are computed for all nodes $i$, following: 
\begin{align}
  lt_{j, i} &= lt_{j-1, 2i} + eq_{j-1, 2i} \times lt_{j-1, 2i+1}
  \label{eq:lt_eq} \\
  eq_{j, i} &= eq_{j-1, 2i} \times eq_{j-1, 2i+1} 
  \label{eq:eq_eq}
\end{align}

The equality comparison is skipped for the first node in every level, and the inequality result of the highest node is returned as the final output of the protocol. Garay et al.~\cite{garay07practical} used arithmetic circuits due to their choice of encryption primitives, such as the use of the Paillier cryptosystem ~\cite{paillier99public} for secure integer multiplications. Later, Blake et al. \cite{blake04strong} showed how integer comparisons could be evaluated using OT. 

\fw{CrypTFlow2}'s algorithm mixes the OT-based integer comparisons of Blake et al. \cite{blake04strong} with the log-depth arithmetic circuit of Garay et al.~\cite{garay07practical} for combining the results of the integer comparisons. It improves the performance of the integer comparisons by folding both the comparisons for one pair, inequality and equality, into one call to $\OT{2^m}{2}$ with a total communication cost of $q(\lambda + 2^{m+1})$ bits for the $q$ pairs. Since the results of the integer comparisons are secret shared bits, a boolean version of the log-depth circuit of Garay et al.~\cite{garay07practical} is implemented by replacing $+$ with bit-XOR ($\oplus$) and $\times$ with bit-AND ($\land$). The $\land$--gates in this boolean circuit are evaluated with a GMW protocol $\prtcl{AND}$ using bit triples. It optimizes bit triple generation with correlated bit triples, observing that for the nodes that output $2$ bits ($j \geq 1$, $i \geq 1$), the $2$ calls to $\prtcl{AND}$ share one operand ($eq_{0,3}$ in Figure~\ref{fig:cf2mill}). The total communication cost to generate the triples is $(\ceil{\log q} - 1)(\lambda + 16) + (q-\ceil{\log q})(\lambda + 8)$, and the total cost to evaluate the boolean circuit is the sum of the triple generation cost and $4(q-1) + 4(q-\ceil{\log q})$ to share the correction bits ($4$ per $\prtcl{AND}$). 

\fw{Cheetah} optimized this by porting the OT primitives to silent OT extension~\cite{kang20ferret} which reduced the communication cost to $q\times(2^{m+1}+m)$ for the $q$ calls to $\OT{2^m}{2}$ and only to $4(q\!-\!1) + 4(q\!-\!\ceil{\log q})$ for the boolean circuit. It follows the strategy presented by Asharov et al.~\cite{asharov13more} and generates a bit triple with 2 calls to $\ROT{2}{1}$. Since these ROTs use silent OT extension, $\ROT{2}{1}$ is almost free in terms of communication. \fw{Cheetah} follows \fw{CrypTFlow2} on setting the segment size $m\!\!=\!\!4$ and using $\OT{16}{2}$ for the best performance. For $m\!\!=\!\!4$, communication requires $9b$ bits for integer comparisons and $2b\!+4\!-4\!\ceil{\log b}$ bits for the boolean circuit, totaling approximately $11b$ bits, while $m\!\!=\!\!1$ requires $5b$ bits for comparisons and $8b-4-4\ceil{\log b}$ bits for circuit evaluation, totaling about $13b$ bits (an $18\%$ increase). In our evaluation with $2^{13}$ comparisons of $32$ bit numbers, while $\OT{2^m}{2}$ operations take $90$ ms for both settings, $\ROT{2}{1}$ operations vary drastically: $110$ ms for $m\!\!=\!\!1$; $15$ ms for $m\!\!=\!\!4$.
%-------------------------------------------------------------------------------
\subsection{Fully-Boolean Algorithm}
\label{sec:boolalg}
%-------------------------------------------------------------------------------
\fw{SecONNds} features a fully-boolean GMW protocol $\prtcl{MILL}$ shown in Algorithm~\ref{algo:mill}, that eliminates LUT evaluations (chosen OTs). For segments of size $m=1$, while prior works employ heavy $\OT{2}{2}$ for the simple task of comparing a pair of bits, \fw{SecONNds} builds on the observation that $\mathbbm{1}\{x_i = y_i\} = (1 \oplus x_i) \oplus y_i$ and $\mathbbm{1}\{x_i < y_i\} = (1 \oplus x_i) \land y_i$. Evaluating the equality comparisons for the bit pairs is free, $\mathcal{P}_0$ sets its share of the equals result $\sshr{eq_{0, i}}{0}{2} = 1 \oplus x_i$ and $\mathcal{P}_1$ sets its share $\sshr{eq_{0, i}}{1}{2} = y_i$, no communication is required. For inequalities, both parties make one call to $\prtcl{AND}$ for each bit where $\mathcal{P}_0$ inputs $1 \oplus x_i$, $\mathcal{P}_1$ inputs $y_i$ and set their shares of $lt_{0, i}$ to the output of $\prtcl{AND}$. This approach only uses $\ROT{2}{1}$ (for triple generation) and takes less than $40$ ms for $2^{13}$ comparisons with $32$ bit numbers. It requires a communication of just $4$ bits per $\prtcl{AND}$ and a total of $4b$ bits. 

For the log-depth circuit, we get a total communication cost of less than $12b$ bits for the millionaires' protocol. Although this is better than $16b$ in the case of using $\OT{2^m}{2}$ for $m=1$, it is still higher than $11b$ in the \emph{optimal setting} of $m=4$. Observe that, alternately to the log-depth strategy, we can also combine the integer comparison results serially in the following manner:
\begin{align}
    lt_i &= lt_{i-1} \oplus lt_{0, i} \oplus (eq_{0, i} \land lt_{0, i-1}) \text{ for } i \in [1, b)
    \label{eq:lt_i}
\end{align}

Following this linear strategy with $lt_0 = 0$, the final result is produced in the value $lt_{q-1}$, and the overall computation requires only $b-1$ calls to $\prtcl{AND}$ which is roughly half of $2 b - 1 - \ceil{\log b}$ in the case of the log-depth strategy with $m=1$. This strategy incurs a communication cost of only $4(b-1)$ and brings the total communication footprint of the millionaires' protocol to under $8b$ bits which is $27\%$ lower than the $11b$ bits cost of the $\OT{16}{2}$ with $m=4$. 

\begin{algorithm}[t]
\setstretch{1.6} % Set line spacing
\caption{$\prtcl{MILL}$ Millionaires' in \fwbold{SecONNds}}
\label{algo:mill}
\DontPrintSemicolon
\SetAlgoVlined
    \KwInput
    {Data bitwidth $b$; Inequality $g\!:\!\{0,1\}\!\rightarrow\!\{ <,> \};$\newline
    Input $i_p$
    }
    \KwOutput{Output secret share $\sshr{o}{p}{2^b}$}

    \For{$i = 0$ \KwTo $b-1$}
    {
        \tcc{\hfill \underline{\textnormal{Bit Extraction \& Share Generation}  }}  
        % $\sshr{\vctr{i_0}}{p}{2}[i] = \left( \left( i_p \gg i\right) \land 1 \right) \oplus g \oplus 1$ \\
        $\sshr{{b_0}}{p}{2} = \Big(\!\big( i_p / 2^i \bmod 2 \big) \oplus g'\Big) \land p' $ \\
        $\sshr{{b_1}}{p}{2} = \Big(\!\big( i_p / 2^i \bmod 2 \big) \oplus g \,\,\Big) \land p $
        
        \tcc{\hfill \underline{\textnormal{Bit Equality \& Inequality Comparisons}}} 

        $\sshr{\vctr{b_{eql}}}{p}{2}[i] = \sshr{{b_0}}{p}{2} \oplus \sshr{{b_1}}{p}{2}$\\
        $\sshr{\vctr{b_{l/g}}}{p}{2}[i] = \prtcl{AND} \Big( \sshr{{b_0}}{p}{2}, \,\sshr{{b_1}}{p}{2} \Big) $
    }
    \tcc{\hfill \underline{\textnormal{Combining Bit Results}}}
    \For{$i = 0$ \KwTo $b-2$}
    {

        $\sshr{b_{and}}{p}{2} = \prtcl{AND} \Big( \sshr{\vctr{b_{eql}}}{p}{2}[i+1], \,\sshr{\vctr{b_{l/g}}}{p}{2}[i] \Big) $
        
        $\sshr{\vctr{b_{l/g}}}{p}{2}[i + 1] = \sshr{\vctr{b_{l/g}}}{p}{2}[i + 1] \oplus \sshr{b_{and}}{p}{2}$
    }

    $\sshr{o}{p}{2} = \sshr{\vctr{b_{l/g}}}{p}{2}[b - 1]$
\end{algorithm}

\begin{table}[t]
\centering
\caption{Costs for Millionaires' Protocol}
\label{tab:mill_costs}
\begin{tabular}{cccc}
\toprule
\textbf{Protocol} & \textbf{Communication} & \textbf{Runtime} \big($2^{13}$ calls\big) \\
\midrule
$\mathsf{Cheetah}$ & \multirow{2}{*}{\( 13b - 4 - 4\lceil\log b\rceil\)} & \multirow{2}{*}{\(\approx 200\) ms} \\
(m = 1) & & \\
\midrule
$\mathsf{Cheetah}$ & \multirow{2}{*}{\( 11b + 4 - 4\lceil\log b\rceil\)} & \multirow{2}{*}{\(\approx 105\) ms} \\
(m = 4) & & \\
\midrule
\fwbold{SecONNds} & \multirow{2}{*}{\(< 8b\)} & \(< \big(70 + \textbf{5}\big)\) ms \\
(ours) & & Offline $+$ \textbf{Online} \\
\bottomrule
\end{tabular}
\end{table}

The linear approach requires half as many calls to $\ROT{2}{1}$ and takes less than $35$ ms for $2^{13}$ comparisons with $32$ bit numbers. In Table~\ref{tab:mill_costs}, we show the total cost of the millionaires' protocol implemented in \fw{SecONNds}, the runtimes are reported for $2^{13}$ runs with $b=32$. The total computation time for our new fully Boolean algorithm for the millionaires' protocol is under $75$ ms with online triple generation, which is $28\%$ less than the $\OT{16}{2}$ version with $m=4$. Note that while the linear strategy halves the communication footprint as well as the computation cost, it incurs an exponential increase in the number of rounds. Application developers can implement a simple toggle to switch between both strategies depending on available network resources to ensure the best quality of service. In the following section, we show how to significantly lower the online runtime from $75$ ms to under $5$ ms with offline triple generation.
%-------------------------------------------------------------------------------
\subsection{Offline Triple Generation}
\label{sec:tripgen}
%-------------------------------------------------------------------------------
To efficiently generate bit triples with silent OT extension and to safely shift the triple generation to the offline phase, we implement an offline triple generator with an internal buffer, inspired by the PRNG implementation in the cryptoTools library~\cite{cryptoTools}. The triple generator automatically generates enough triples in chunks of fixed size to fill its buffer, as soon as a network connection with a user is established. When a query is requested, all underlying protocols make use of the \texttt{get} functionality of the triple generator to access the preprocessed triples. The triple generator automatically generates new triples and refills the buffer when it is exhausted during the online computation. In case a protocol requests a volume of triples larger than the buffer size, the buffer size is incremented, and the generator generates new triples in chunks of fixed size to fill the buffer. 

Chunking increases the complexity of the communication for $n$ calls to $\ROT{2}{1}$ from $O\left(\log n \right)$ to a sublinear $O\left(\frac{n}{m} \log m\right)$ where $m$ is the size of one chunk. For any large $m$, the communication footprint of our chunking strategy is fairly comparable to the naive approach of generating $n$ $\ROT{2}{1}$'s in one shot. The real advantage of the chunking strategy comes in terms of computation time, which is actually the main concern with silent OT. The computational complexity involved in naively generating $n$ $\ROT{2}{1}$'s is $O\left(n^2 \right)$, arising from the matrix multiplication involved in the LPN encoding phase of silent OT. With the chunking strategy, the computation complexity is significantly reduced to $O\left( nm \right)$ which is now linear in $n$.
%-------------------------------------------------------------------------------
%-------------------------------------------------------------------------------
\section{Nonlinear Operations}
\label{sec:nonlin}
%-------------------------------------------------------------------------------
%-------------------------------------------------------------------------------
In this section, we review the ReLU and Truncation protocols in \fw{SecONNds} for quantized CNNs. Max Pooling and Average Pooling protocols are discussed in Appendixes~\ref{sec:maxpool} and ~\ref{sec:avgpool}.
%-------------------------------------------------------------------------------
\subsection{ReLU}
\label{sec:relu}
%-------------------------------------------------------------------------------
The function $\mathsf{ReLU}$ (Rectified Linear Unit) is a widely used activation function in neural networks, defined as $\mathsf{ReLU}(i) = \max(0, i)$. In \fw{SecONNds}, we employ \fw{Cheetah}'s Silent OT-based implementation of the \fw{CrypTFlow2} protocol with the millionaires' protocol described in Section~\ref{sec:mill}. This protocol, denoted as $\prtcl{ReLU}$, is shown in Algorithm~\ref{algo:relu}.

\begin{algorithm}[t]
\setstretch{1.6}
\caption{$\prtcl{ReLU}$ ReLU}
\label{algo:relu}
\DontPrintSemicolon
    \KwInput
    {Input secret share $\sshr{i}{p}{2^b}$}
    \KwOutput{Output secret share $\sshr{o}{p}{2^b}$}

    $\mathsf{MSB}\left(\sshr{i}{p}{2^b} \right) = \sshr{i}{p}{2^b}/2^{b - 1} $
    
    $\left|\sshr{i}{p}{2^b}\right| = \sshr{i}{p}{2^b} - \mathsf{MSB}\!\left(\sshr{i}{p}{2^b}\right) \cdot 2^{b - 1}$

    $i_{mill} = {(-1)}^p\left|\sshr{i}{p}{2^b}\right| + p \cdot \left( 2^{b - 1} - 1 \right)$ 

    $\sshr{w}{p}{2} = \prtcl{MILL}\big(b-1, 1 , i_{mill}\big)$

    $\sshr{i_{drelu}}{p}{2} = \mathsf{MSB}\!\left(\sshr{i}{p}{2^b}\right) \oplus \sshr{w}{p}{2} \oplus p'$\tcp*[r]{\textnormal{\!\!dReLU result}}

    $\delta = \left( 1 - 2 \cdot \sshr{i_{drelu}}{p}{2} \right)\!\cdot\!\sshr{i}{p}{2^b} $  \tcp*[r]{\textnormal{Delta for COT}}
    $c = \sshr{i_{drelu}}{p}{2}$ \tcp*[r]{\textnormal{Choice for COT}}
    
    $m_s = \COTsnd{2}{b}(\delta)$\quad;\quad$m_r = \COTrec{2}{b}(c)$ \\
    $\sshr{o}{p}{2^b} = \Big( \sshr{i}{p}{2^b} \cdot \sshr{i_{drelu}}{p}{2} + m_r - m_s \Big) \bmod 2^b$
\end{algorithm}

The protocol takes as input the secret shares of the activations entering the $\mathsf{ReLU}$ layer and returns the secret shares of the $\mathsf{ReLU}$ result. The protocol first evaluates $d\mathsf{ReLU} = \mathbbm{1}\{i > 0\}$ and returns fresh secret shares of $i$ if $d\mathsf{ReLU}$ is $1$ and secret shares of $0$ if $d\mathsf{ReLU}$ is zero. Observe that $i > 0$ in $\mathbb{Z}_{2^b}$ corresponds to $i < 2^{b-1}$, which is equivalent to:
\begin{align}
\bigg(\!\mathsf{MSB}\Big(\!\sshr{i}{0}{2^b}\!\Big)\!+\!\mathsf{MSB}\Big(\!\sshr{i}{1}{2^b}\!\Big)\!\!\bigg)\!\cdot\!2^{b\!-\!1} + \left|\sshr{i}{0}{2^b}\!\right| + \left|\sshr{i}{1}{2^b}\!\right| < 2^{b\!-\!1} \nonumber
\end{align}

\noindent This inequality depends only on the sum of the absolute values of the shares wrapping around the maximum absolute value in the ring, $2^{b-1}-1$, denoted by the bit $w$ in Algorithm~\ref{algo:relu}, and the equality of the most significant bits (MSB) of the shares. Particularly, it holds only if $w$ is $0$ and both MSBs are the same, or if $w$ is $1$ and both MSBs are different:
\begin{align}
d\mathsf{ReLU} &= \mathsf{MSB} \Big( \sshr{i}{0}{2^b} \Big) \oplus \mathsf{MSB} \Big( \sshr{i}{1}{2^b} \Big) \oplus \sshr{w}{0}{2} \oplus \sshr{w}{1}{2}  \oplus 1 \nonumber
\end{align}

\noindent Here, the wrap bit $w$ is securely computed using the millionaires' protocol described in Section~\ref{sec:mill}. After computing $d\mathsf{ReLU}$, the protocol uses a secure multiplexer functionality, $\mathsf{MUX}$, realized with two calls to $\COT{2}{b}$. The $d\mathsf{ReLU}$ result serves as the selection bit. If the $d\mathsf{ReLU}$ result is $1$, the $\mathsf{MUX}$ outputs new shares of the input; if the $d\mathsf{ReLU}$ result is $0$, the $\mathsf{MUX}$ outputs shares of zero.

\begin{algorithm}[t]
\setstretch{1.6} % Set line spacing
\caption{$\prtcl{Trunc}$ Truncation}
\label{algo:truncate}
\DontPrintSemicolon
\SetAlgoVlined
    \KwInput
    {Right shift amount $s$ ;\newline
    Bit $i_{msb}$ indicating if $\mathsf{MSB}(i)$ is known; \newline
    Input secret share $\sshr{i}{p}{2^b}$ where $i<2^{b-1}$
    }
    \KwOutput{Output secret share $\sshr{o}{p}{2^b}$}
    
    $\mathsf{MSB}\left(\sshr{i}{p}{2^b} \right) = \sshr{i}{p}{2^b}/2^{b - 1} $\\    \tcc{\hfill \underline{\textnormal{Wrap Bit Computation}}}
    \If{$i_{msb}$}
    {
        $\sshr{w}{p}{2} = \prtcl{AND}\!\bigg(p\!\cdot\!\mathsf{MSB}\Big(\!\sshr{i}{p}{2^b}\!\Big),\,  p'\!\cdot\!\mathsf{MSB}\Big(\!\sshr{i}{p}{2^b}\!\Big)\!\bigg)$\\
    }
    \Else
    {
        $\sshr{w}{p}{2} = \prtcl{MILL}\bigg(b, 1 , \Big[{(-1)}^p\sshr{i}{p}{2^b} + p \cdot \left( 2^{b} - 1 \right)\Big]\bigg)$
    }
    \tcc{\hfill \underline{\textnormal{Wrap B2A Share Conversion}}}
    \If{$p = 0$}
    {
        $\delta = -2 \cdot \sshr{w}{p}{2}$ \tcp*[r]{\textnormal{Delta for COT}}
        $m_s = \COTsnd{2}{b}(\delta)$ \\
        $\sshr{w}{p}{2^b} = \sshr{w}{p}{2} - m_s$
    }
    \Else
    {
        $c = \sshr{w}{p}{2}$ \tcp*[r]{\textnormal{Choice for COT}}
        $m_r = \COTrec{2}{b}(c)$ \\
        $\sshr{w}{p}{2^b} = \sshr{w}{p}{2} + m_r$
    }
    \tcc{\hfill \underline{\textnormal{Final Truncation Result}}}
    $\sshr{o}{p}{2^b} =  \sshr{i}{p}{2^b} / 2^s  - \sshr{w}{p}{2^b} \cdot 2^{b-s}$
    
\end{algorithm}
%-------------------------------------------------------------------------------
\subsection{Truncation}
\label{sec:trunc}
%-------------------------------------------------------------------------------
In the context of fixed-point computation, truncation is a crucial operation to prevent the data scale from escalating after multiplications. We present the protocol employed in \fw{SecONNds}, denoted as $\prtcl{Trunc}$, in Algorithm~\ref{algo:truncate}. The core concept of this approach is to represent the data in its unsigned form as follows:

\begin{align}
    w &= \mathbbm{1}\Big\{ \sshr{i}{0}{2^b} + \sshr{i}{1}{2^b} > 2^b - 1 \Big\} \nonumber \\
    i &= \sshr{i}{0}{2^b} + \sshr{i}{1}{2^b} - w \cdot 2^b \nonumber \\
    i/2^s &\approx \sshr{i}{0}{2^b}/2^s + \sshr{i}{1}{2^b}/2^s - w \cdot 2^{b-s} \nonumber
\end{align}

This computation is approximate, as it does not account for potential carries from the dropped portion of the secret shares. However, this introduces an error only in the least significant bit of the result, and previous work~\cite{secureQ8, huang22cheetah} has shown that neural networks are highly tolerant to this particular $1$ bit error in truncation, with negligible impact on performance. The wrap bit $w$ is computed using $\prtcl{MILL}$, but when the sign of the secret value is known, such as post-ReLU when the values are positive, $w = \mathsf{MSB} \Big( \sshr{i}{0}{2^b} \Big) \land \mathsf{MSB} \Big( \sshr{i}{1}{2^b}\Big)$ and can be computed with a single call to $\prtcl{AND}$, both $\prtcl{MILL}$ and $\prtcl{AND}$ use offline bit triples. In total, the main secure computation operations in this protocol are $2b-1$ calls to $\prtcl{AND}$ for $\prtcl{MILL}$ and one call to $\COT{2}{b}$ to convert the secret shares of $w$ from binary to arithmetic, or just one $\prtcl{AND}$ and one $\COT{2}{b}$ if the MSB of the secret share is known.

\subsection{Secret-sharing in $\mathbb{Z}_{2^b}$ vs. $\mathbb{Z}_{P}$}
\label{sec:2vP}

The choice of the ring for secret sharing, $\mathbb{Z}_{2^b}$ (a power of two) or $\mathbb{Z}_{P}$ (where $P$ is prime), significantly influences the implementation and performance of nonlinear operations. In $\mathbb{Z}_{P}$, any protocol requiring a comparison (like wrap bit computation) must also check if the secret overflows $P$ and must make provision for handling it. On the other hand, operations in $\mathbb{Z}_{2^b}$ benefit from natural alignment with binary systems; the boolean circuit in $\prtcl{MILL}$ also handles only an explicitly specified bitwidth and overflows naturally wrap around.

\begin{figure*}[h]
    \centering
    \includegraphics[width=\textwidth]{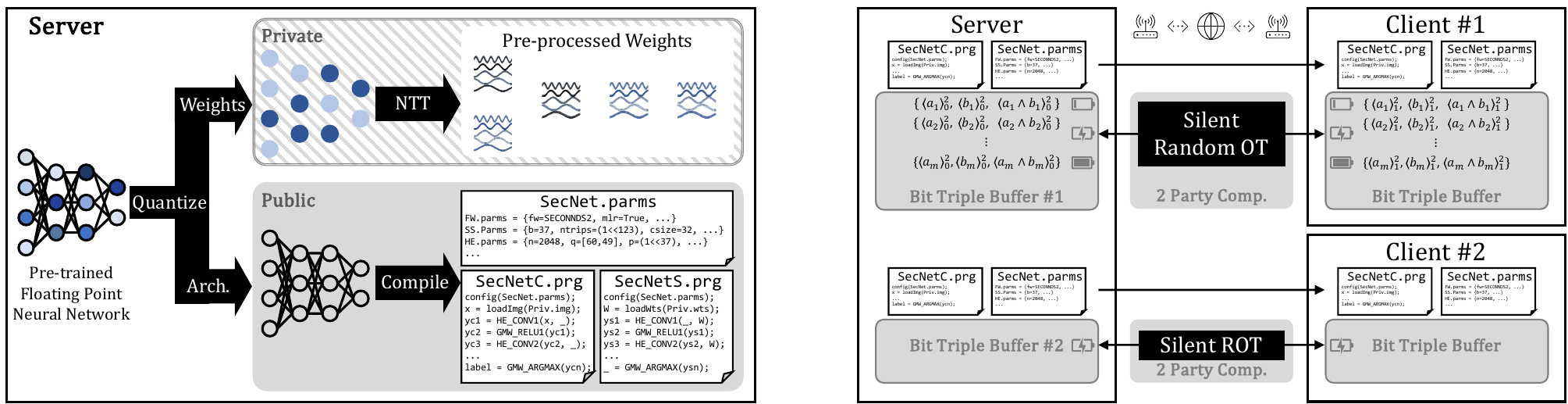}
    \caption{Workflow of server (left) and client (right) setup in \fw{SecONNds}.}
    \label{fig:setup}
\end{figure*}
%-------------------------------------------------------------------------------
%-------------------------------------------------------------------------------
\section{Linear Operations}
\label{sec:linear}
%-------------------------------------------------------------------------------
%-------------------------------------------------------------------------------
The typical linear layers in CNNs consist of the convolution layers, fully-connected/matrix multiplication layers, and batch normalization. The bulk of linear operations comes from convolutions with matrix multiplications that appear only at the very end of the CNN. Batch normalization typically appears after convolutions, and it is common practice to fuse batch normalization with convolution~\cite{markus2023fusing} during inference.
%-------------------------------------------------------------------------------
\subsection{HE Kernels}
\label{sec:hekernels}
%-------------------------------------------------------------------------------
Linear algebra operations can be composed from vector multiplications, rotations, and additions, all of which are supported by modern HE schemes for the plaintext space $Z_P^N$ where $P$ is the HE plaintext modulus prime. \fw{HELiKs}~\cite{balla23heliks} offers state-of-the-art kernels that compose these operations in the most efficient manner. At the core of these protocols is an iterative algorithm where each iteration involves a HE multiplication of the input vector with the weights, accumulating the product into the previous iteration's result, and finally a HE rotation to shift the accumulated value to adjust for the next iteration. This strategy significantly reduces the number of HE rotations required. \fw{HELiKs} further boosts the performance by preprocessing the weights with NTT which leads to a very fast online runtime. Although \fw{HELiKs} offers cutting-edge performance for linear algebra operations, for secure inference, it necessitates the use of $Z_P$ for nonlinear operations or the use of a share conversion protocol to convert secret shares from $Z_P$ to secret shares in $Z_{2^b}$. 

\fw{Cheetah}'s HE kernels operate in the plaintext space $Z_{2^b}^N$. It encodes secret data into the polynomial coefficients, enabling polynomial multiplications and additions to secret data through HE multiplications and additions. \fw{Cheetah}'s HE kernels compute the linear algebra operations purely through iterative HE multiply-accumulate (MAC) operations without any HE rotations. They produce sparse results in HE, and then extract just the relevant coefficients from these results to produce the final results for the linear algebra operation. 

%-------------------------------------------------------------------------------
\subsection{NTT preprocessing}
\label{sec:nttpreproc}
%-------------------------------------------------------------------------------
We optimize \fw{Cheetah}'s HE kernels for $Z_{2^b}^N$ with NTT-preprocessing. HE multiplications involve polynomial multiplication, which is $O(N^2)$ in the computational complexity. The polynomial degree modulus $N$ is typically of the order of thousands or higher and induces a very long runtime for multiplication. HE libraries typically optimize this operation by transforming both operands with NTT, performing a Hadamard product on the transformed operands, and transforming the product back to its natural representation with iNTT (inverse NTT). Every HE multiplication involves two NTT operations and one iNTT operation, both involving a complexity of $O(N \log N)$. 

Balla et al.~\cite{balla23heliks} observed that, for linear algebra operations, most of the calls to HE multiplication in the same query reuse one of the operands (input vector) and the other operand (weights) is reused over different queries and known to the server (computing party). In \fw{SecONNds}, the weights are automatically preprocessed with NTT during encoding and are always maintained in the NTT representations. During the online query, on receipt of the input ciphertexts, the server first transforms each ciphertext with NTT, performs all HE MAC operations in NTT, and only transforms the final HE results back from NTT. Since these HE results are sparse, only coefficients that contain the elements of the output vector are extracted and sent back to the client for decryption. 
%-------------------------------------------------------------------------------
\subsection{GPU Acceleration}
\label{sec:gpuacc}
%-------------------------------------------------------------------------------
HE operations are local and non-interactive, making them a prime candidate for hardware acceleration. The one-ended computation of HE protocols requires just the computing party to have access to the hardware accelerator. In remote cloud computation, it is very common for servers to possess GPUs. Also, polynomial data types are represented with vectors, which are perfect for GPU SIMD computation.

\emph{Troy}~\cite{troy} is a new software library that implements the SEAL HE Library~\cite{SEALv4} in CUDA~\cite{cuda}. \fw{SecONNds} employs the HE evaluator from Troy for GPU implementations of the server's HE computation. All of the client's computation is performed with the standard SEAL Library; the new GPU implementations do not handle any secret key related operations and purely compute only on already encrypted data.
%-------------------------------------------------------------------------------
%-------------------------------------------------------------------------------
\section{Framework Overview}
\label{sec:overview}
%-------------------------------------------------------------------------------
%-------------------------------------------------------------------------------

\begin{table*}[t]
\centering
\caption{Operation Runtimes (in seconds) and Communication (in MiB) for Nonlinear operations}
\label{table:nonlin}
\resizebox{\textwidth}{!}{
\begin{tabular}{ccccccccccccccc}
\toprule
\textbf{Nonlinear} & \multicolumn{2}{c}{\fw{CrypTFlow2}} & \multicolumn{2}{c}{\fw{HELiKs}} & \multicolumn{2}{c}{\fw{Cheetah}} & \multicolumn{2}{c}{\fw{SecNN\!\!-\!\!P} (LR)} & \multicolumn{2}{c}{\fw{SecONNds\text{-}P}} & \multicolumn{2}{c}{\fw{SecNN} (LR)} & \multicolumn{2}{c}{\fwbold{SecONNds}} \\
\textbf{Operation}& Time   & \emph{Comm.}& Time & \emph{Comm.}& Time & \emph{Comm.}& Time & \emph{Comm.}& Time   & \emph{Comm.}& Time     & \emph{Comm.}& Time  & \emph{Comm.}\\
\midrule
Truncation        & $4.65$   & $\mathit{307}$  & $4.65$ & $\mathit{307}$  & $0.12$ & $\mathit{4.86}$ & $0.81$ & $\mathit{284}$  & $0.51$   & $\mathit{267}$  & $0.07$     & $\mathit{4.79}$ & $0.06$  & $\mathit{4.62}$ \\
ReLU              & $6.73$   & $\mathit{302}$  & $6.73$ & $\mathit{302}$  & $2.53$ & $\mathit{110}$  & $1.72$ & $\mathit{246}$  & $0.87$   & $\mathit{186}$  & $0.72$     & $\mathit{116}$  & $0.23$  & $\mathit{88.7}$ \\
Max Pooling       & $7.10$   & $\mathit{398}$  & $7.10$ & $\mathit{398}$  & $4.34$ & $\mathit{154}$  & $2.04$ & $\mathit{326}$  & $1.03$   & $\mathit{247}$  & $0.89$     & $\mathit{151}$  & $0.26$  & $\mathit{117}$  \\
Avg Pooling       & $0.03$   & $\mathit{4.19}$ & $0.03$ & $\mathit{4.23}$ & $0.06$ & $\mathit{4.43}$ & $0.02$ & $\mathit{4.20}$ & $0.02$   & $\mathit{4.19}$ & $0.06$     & $\mathit{4.42}$ & $0.05$  & $\mathit{4.45}$ \\
Arg Max           & $0.06$   & $\mathit{0.21}$ & $0.06$ & $\mathit{0.19}$ & $0.02$ & $\mathit{0.11}$ & $0.02$ & $\mathit{0.19}$ & $0.02$   & $\mathit{0.20}$ & $0.01$     & $\mathit{0.11}$ & $0.01$  & $\mathit{0.11}$ \\
\bottomrule
\end{tabular}
}
\end{table*}

\begin{table}[t]
\caption{Offline Triple Generation costs}
\label{table:tripgen}
\centering
\resizebox{\columnwidth}{!}{%
\begin{tabular}{cccc}
\toprule
\textbf{Framework}&\textbf{Triples}     &\textbf{Runtime (s)}  &\textbf{Comm. (MiB)} \\
\midrule
\fw{SecONNds\text{-}P} (LR)   &\(1.2\!\times\!10^9\)&\(27.16\)&\(60.41\)\\
\fw{SecONNds\text{-}P}         &\(8.2\!\times\!10^8\)&\(22.43\)&\(43.12\)\\
\fw{SecONNds} (LR)        &\(4.9\!\times\!10^8\)&\(10.61\)&\(26.36\)\\
\fwbold{SecONNds}          &\(3.4\!\times\!10^8\)&\(8.31\) &\(17.01\)\\
\bottomrule
\end{tabular}
}
\end{table}

\fw{SecONNds} follows a modular design with distinct setup phases for server and client, illustrated in Figure~\ref{fig:setup}. The server's setup involves three key steps: (1) quantizing the pre-trained model to fixed-point representation, packing and encoding weights into HE plaintexts with NTT transformations for efficient polynomial multiplication, (2) compiling the network architecture into configuration files specifying protocol parameters including secret sharing bitwidth, $\prtcl{MILL}$ variant (log-depth/linear), NTT preprocessing settings for convolutions, and triple buffer configurations, and (3) generating program files containing the complete computation graph and layer-wise execution order for both server (\texttt{SecNetS.prg}) and client (\texttt{SecNetC.prg}) implementations. This preprocessing phase is query-independent and needs to be performed only once unless the model parameters are updated.

A client requesting secure inference service first receives the model-specific configuration and program files. Based on these specifications, the client generates the necessary number of Beaver's bit triples using silent ROT, determined by the model architecture and planned number of inferences. The triple generation process, being input-independent, is entirely preprocessing and can be performed offline before the actual inference requests. These triples are stored in a buffer that automatically refills when exhausted during computation, with dynamic size adjustment capabilities to handle varying protocol requirements.

The online inference protocol executes layer-by-layer with both parties maintaining secret shares of intermediate activations throughout the network. Nonlinear operations (ReLU, Max Pooling) employ GMW protocol with the preprocessed triples, requiring interaction only for AND gates where parties exchange correction bits. For linear operations (Convolution, Fully Connected layers), the client encrypts and sends its activation shares to the server, which adds its own shares to these ciphertexts, performs linear operations using the NTT-preprocessed weights, applies a random mask for security, and returns the encrypted result to the client for decryption into output shares. This mixed-protocol approach optimally balances computation and communication overhead.

\fw{SecONNds} ensures perfect security under the semi-honest model - the client learns only the final classification label while the server learns nothing about the input or intermediate values. Its modular design allows for runtime protocol selection and parameter configuration through the configuration files, enabling optimization for different network conditions and performance requirements. For example, developers can toggle between log-depth and linear variants of $\prtcl{MILL}$ based on network latency, or enable/disable NTT preprocessing for convolutions depending on available computational resources.
%-------------------------------------------------------------------------------
%-------------------------------------------------------------------------------
\section{Evaluation}
\label{sec:eval}
%-------------------------------------------------------------------------------
%-------------------------------------------------------------------------------

\fw{SecONNds} is implemented in the OpenCheetah~\cite{OpenCheetah2024} variant of the Secure and Correct Inference (SCI) Library~\cite{SCI2024}. We implement \fw{SecONNds} for secret-sharing with $\mathbb{Z}_{2^b}$ and \fw{SecONNds\text{-}P} for $\mathbb{Z}_P$, where $P$ is a BFV-SIMD plaintext prime modulus. \fw{SecONNds} uses $1$-bit approximate truncation, while \fw{SecONNds\text{-}P} employs faithful truncation and returns bit-exact results compared to the plaintext model. Both use the fast version of $\prtcl{MILL}$ with the linear strategy by default. If the low-round variant is employed for a framework, it is denoted by LR in parentheses, e.g., \fw{SecONNds} (LR).

For all our evaluations, we outfitted all frameworks being compared with: Ferret Silent OT Extension~\cite{kang20ferret} from the EMP-OT Library~\cite{EMPOT2024}; BFV HE scheme~\cite{brakerski11bfv, fan12bfv} from the SEAL Library~\cite{SEALv4}; and server-side GPU acceleration with the HE Evaluator from the Troy Library~\cite{troy}. All CPU operations of both parties were performed on $16$ threads of an \emph{Intel Xeon Gold 6338} processor, supplemented by $1$ TB of RAM and utilizing both the \emph{AES-NI} and \emph{AVX-512} instruction set extensions. The server-side GPU evaluations were performed on an \emph{NVIDIA RTX A6000} system.

We evaluate the performance of \fw{SecONNds} on the pre-trained SqueezeNet~\cite{iandola16squeezenet} CNN model from OpenCheetah~\cite{OpenCheetah2024}. The SqueezeNet model is uniformly quantized to fixed-point with a bitwidth of $37$ and $12$ bits for scale, achieving $79.6\%$ Top-5 accuracy, the same as in prior works. \fw{SecONNds\text{-}P} performs bit-exact computations and produces the same output logits as the cleartext model. \fw{SecONNds} uses $1$-bit approximate truncation; with just \emph{$\mathit{0.0015\%}$ Mean Absolute Percentage Error} (MAPE) in output logits, it bears no impact on accuracy. We also evaluate a ResNet50 model~\cite{he16resnet}, which achieves $92.3\%$ Top-5 accuracy with the same $37$-bit setup and $12$-bit scale and present the results in Appendix~\ref{sec:resnet}.

%-------------------------------------------------------------------------------
\subsection{Offline Preprocessing}
\label{sec:offline}
%-------------------------------------------------------------------------------
\fw{SecONNds} performs NTT preprocessing on the model weights for fast online HE computation. This process does not require any secret key information and, moreover, it is completely query independent; no communication is required. For SqueezeNet, \fw{SecONNds} requires $0.73$ seconds for NTT preprocessing, while \fw{SecONNds\text{-}P} requires $2.62$ seconds with HE operations in $\mathbb{Z}_P$ from \fw{HELiKs}. The server can reuse the NTT preprocessed weights over multiple queries unless the model parameters are updated or modified.

\fw{SecONNds} also generates Beaver's triples offline for each query, which significantly improves the online performance of nonlinear operations. \fw{SecONNds\text{-}P} requires $2.4\times$ more triples compared to \fw{SecONNds}, due to the higher overhead associated with nonlinear operations in $\mathbb{Z}_P$. As mentioned in Section~\ref{sec:boolalg}, the log-depth variant of the $\prtcl{MILL}$ (LR) involves up to $2\times$ more AND ($\land$) gates. This leads to a larger volume of communication in each round, resulting in a total communication increase of approximately $1.5\times$ for all calls to $\prtcl{MILL}$ in nonlinear operations, and longer runtimes by a factor of $1.6\times$, when compared to the faster $\prtcl{MILL}$ with the linear circuit.
\begin{figure*}[t]
    \centering
    \includegraphics[width=\textwidth]{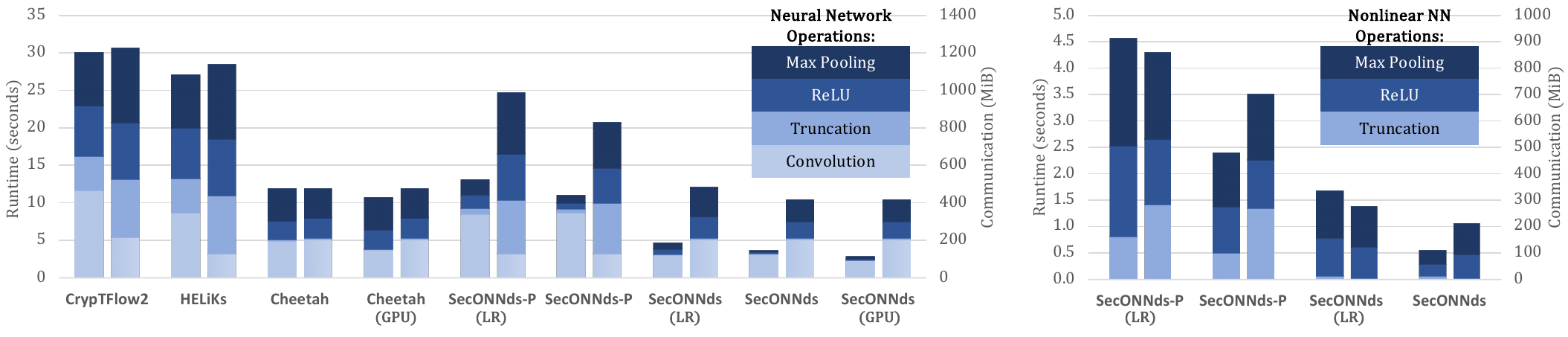}
    \caption{End-to-end (E2E) runtime (left bar) and communication (right bar) performance of each framework for the Neural Network (NN) operations in SqueezeNet, and a close up of nonlinear NN operations for variants of \fw{SecONNds} on the right.}
    \label{fig:e2e_all}
\end{figure*}
%-------------------------------------------------------------------------------
\subsection{Nonlinear layers}
\label{sec:nonlineval}
%-------------------------------------------------------------------------------
In Table~\ref{table:nonlin}, we show the cumulative performance of all nonlinear operations in a single secure inference on SqueezeNet: ReLU, Max Pooling, Average Pooling, ArgMax, and Truncation. \fw{SecONNds} achieves substantial improvements over \fw{Cheetah} in nonlinear operations, with $17\times$ faster Max Pooling, $11\times$ faster ReLU, and $2\times$ faster Truncation. Additionally, communication costs are reduced by up to $20\%$, with most gains observed in ReLU and Max Pooling.

Compared to \fw{CrypTFlow2} and \fw{HELiKs}, \fw{SecONNds\text{-}P} achieves significant runtime reductions -- $8\times$ faster for ReLU, $7\times$ faster for Max Pooling, and $9\times$ faster for Truncation. Communication costs of \fw{SecONNds\text{-}P} are lower by approximately $27\%$ relative to these frameworks. The LR versions of \fw{SecONNds} and \fw{SecONNds\text{-}P} demonstrate increased runtimes and communication volumes due to the added $\land$-gates, and still offer competitive performance, making them suitable when balancing latency and communication for different round complexities under different network conditions.
%-------------------------------------------------------------------------------
\subsection{Linear Layers}
\label{sec:lineval}
%-------------------------------------------------------------------------------
\begin{table}[t]
\caption{Costs for HE Convolution}
\label{table:conv}
\centering
\resizebox{\columnwidth}{!}{
\begin{tabular}{cccc}
\toprule
\textbf{Framework}&\textbf{Offline (s)}    &\textbf{Online (s)} &\textbf{Comm. (MiB)} \\
\midrule
\fw{CrypTFlow2}            & $0.00$    & $11.59$  & $217.10$ \\
\fw{HELiKs}                & $2.51$    & $8.62$   & $129.18$ \\
\fw{Cheetah}               & $0.00$    & $4.94$   & $204.84$ \\
\fw{Cheetah} (GPU)         & $0.00$    & $3.72$   & $204.84$ \\
\fw{SecONNds\text{-}P}/(LR)& $2.62$    & $8.47$   & $129.18$ \\
\fwbold{SecONNds}/(LR)     & $0.76$    & $3.09$   & $204.84$ \\
\fwbold{SecONNds} (GPU)    & $0.72$    & $2.26$   & $204.84$ \\
\bottomrule
\end{tabular}
}
\end{table}

The SqueezeNet model only consists of convolutions for linear layers, and was evaluated for both CPU and GPU execution. In Table~\ref{table:conv}, we show the performance of each framework for all convolutions in the model. \fw{SecONNds} demonstrates reduced online runtime compared to \fw{Cheetah} due to NTT preprocessing. On CPU, \fw{SecONNds} achieves a runtime of $3.09$ seconds, a $1.6\times$ improvement over \fw{Cheetah}'s $4.94$ seconds. With GPU, \fw{SecONNds} reduces runtime to $2.26$ seconds, achieving a speedup of $1.4\times$ over \fw{Cheetah} (GPU). In terms of communication, \fw{SecONNds} achieves the same performance as \fw{Cheetah} for the convolution layers, with both requiring $204.84$ MiB . However, \fw{HELiks} and \fw{SecONNds\text{-}P} show the best communication efficiency, requiring only $130$ MiB, benefiting from improved noise management in HE operations.
%-------------------------------------------------------------------------------
\subsection{E2E Evaluation}
\label{sec:e2e}
%-------------------------------------------------------------------------------
In the End-to-End (E2E) evaluations, shown in Figure~\ref{fig:e2e_all}, \fw{SecONNds} demonstrates the best performance in terms of both runtime and communication efficiency, achieving total runtimes of $3.70$ seconds on CPU and $2.87$ seconds on GPU. This is a significant improvement over other frameworks, with a $3.24\times$ speedup compared to \fw{Cheetah} on CPU, which has an online runtime of $12$ seconds. Even with GPU, \fw{SecONNds} exhibits a speedup of approximately $3.8\times$ compared to \fw{Cheetah}'s runtime of $10.79$ seconds. For communication, \fw{SecONNds} incurs a total of $420$ MiB, showing a reduction of approximately $12\%$ compared to \fw{Cheetah} ($478.93$ MiB). \fw{SecONNds\text{-}P}, which ensures bit-exact accuracy with full-precision truncation, demonstrates superior performance compared to \fw{HELiKs} in both runtime and communication metrics. \fw{SecONNds\text{-}P} achieves an online runtime of $11.06$ seconds, which is a $2.46\times$ speedup over \fw{HELiKs}, which has an online runtime of $27.20$ seconds. In terms of communication, \fw{SecONNds\text{-}P} achieves a total of $834.17$ MiB, a $27\%$ reduction from $1141.95$ MiB required by \fw{HELiKs}.

For the logarithmic-depth (LR) variants, both \fw{SecONNds} and \fw{SecONNds\text{-}P} are designed to minimize the number of communication rounds at the cost of slightly increased computational and communication volume. Specifically, \fw{SecONNds} (LR) achieves reduced communication rounds of $1084$, while \fw{SecONNds\text{-}P} (LR) operates with $1542$ rounds. These figures represent a significant reduction compared to the non-LR versions of \fw{SecONNds} and \fw{SecONNds\text{-}P}, which require $4630$ and $5800$ rounds, respectively. Compared to \fw{Cheetah}, which uses $900$ rounds, the LR variants demonstrate competitive performance with a trade-off involving much higher computational effort and communication volume per round. 
%-------------------------------------------------------------------------------
%-------------------------------------------------------------------------------
\section{Analysis and Future Outlook}
\label{sec:future}
%-------------------------------------------------------------------------------
%-------------------------------------------------------------------------------
\subsection{System Limitations} 
%-------------------------------------------------------------------------------
\fw{SecONNds} operates under the semi-honest security model, assuming honest protocol adherence, which restricts its use in malicious adversarial settings without incurring additional overhead. Additionally, secure inference still introduces considerable computational overhead compared to plaintext inference, which still poses challenges for real-time resource-constrained applications. Although the GPU library \emph{Troy}~\cite{troy} significantly accelerates elemental homomorphic encryption operations, the performance improvements for higher-level composite functions like convolutions are modest compared to highly optimized multithreaded CPU implementations. This highlights the need for specialized hardware acceleration targeting composite operations (like linear algebra kernels) for comprehensive performance gains.
%-------------------------------------------------------------------------------
\subsection{Alternate Solutions} 
%-------------------------------------------------------------------------------
In comparison to HE-only frameworks, which achieve secure inference with minimal interaction but suffer from substantial computational overhead due to FHE bootstrapping~\cite{gentry09fhe}, \fw{SecONNds} provides a more balanced solution by reducing these costs. Trusted Execution Environments (TEEs) like Intel SGX~\cite{mckeen13sgx} offer low-latency inference within secure enclaves but rely on hardware trust assumptions and are vulnerable to side-channel attacks~\cite{van18foreshadow, chen19sgxpectre}. \fw{SecONNds} leverages applied cryptography and a mixed-protocol strategy that optimizes both linear and nonlinear operations, achieving practical performance without any dependence on trusted hardware.
%-------------------------------------------------------------------------------
\subsection{Areas for Innovation} 
%-------------------------------------------------------------------------------
Potential enhancements for \fw{SecONNds} could include integrating neural network compression techniques such as pruning~\cite{han15pruning}, knowledge distillation~\cite{hinton2015distilling}, or newer approaches to reduce model complexity and improve efficiency. Expanding on the foundational protocols of \fw{SecONNds} to higher-level ones, say splines for the secure evaluation of complex nonlinear functions like GeLU~\cite{hendrycks2016gaussian} and Swish~\cite{ramachandran2018searching}, would enable support for a broader range of neural network architectures. Additionally, extending the security model to accommodate malicious adversaries and further optimizing hardware acceleration for composite operations could enhance both the security and performance of secure inference tasks.
%-------------------------------------------------------------------------------
%-------------------------------------------------------------------------------
\section{Conclusion}
\label{sec:conclusion}
%-------------------------------------------------------------------------------
%-------------------------------------------------------------------------------
\fw{SecONNds} advances secure neural network inference by addressing key performance bottlenecks and demonstrating practical applicability on large-scale datasets. It is ideal for enabling privacy-sensitive ML applications in resource-constrained environments. It allows service providers to securely outsource NN computation, delivering practical performance with robust security and reduced computational load for users. Moreover, being a non-intrusive framework, it offers foundational modules that apply to any neural network, regardless of the media type, without requiring any model fine-tuning, and cuts any reliance on training data. \fw{SecONNds} is highly compatible -- it is fully open-source, modular and dynamic, allowing for mixing between different preprocesses and protocol optimizations at runtime, e.g., linear layers with online/offline/no NTT preprocessing, toggling round complexity of Millionaires' protocol, etc.

%-------------------------------------------------------------------------------
%-------------------------------------------------------------------------------
\section*{Ethics considerations}
%-------------------------------------------------------------------------------
%-------------------------------------------------------------------------------
The research adheres to ethical guidelines by ensuring privacy preservation during outsourced neural network inference. The cryptographic techniques employed guarantee that neither the client nor the server can access each other's private data, including inputs, intermediate values, or model parameters. The semi-honest security model excludes malicious adversaries, but potential misuse scenarios are minimized through strict adherence to secure computation protocols. The work does not involve human subjects or datasets that would raise additional ethical concerns.
% -------------------------------------------------------------------------------
% -------------------------------------------------------------------------------
\section*{Open Science}
%-------------------------------------------------------------------------------
%-------------------------------------------------------------------------------
The code and implementation details of \fw{SecONNds} are made openly available on GitHub at \url{https://github.com/SecONNds/SecONNds_1_25}. This ensures transparency and reproducibility of the results presented in the paper. All data used, the ImageNet samples, and the pre-trained SqueezeNet model are public resources from prior works that are also directly available from the aforementioned GitHub repository. The protocol procedures are thoroughly described in the paper, and the source code is furnished with detailed comments to facilitate replication. To support further research, detailed configuration files and instructions are provided for deploying the framework in various scenarios.
%-------------------------------------------------------------------------------
%-------------------------------------------------------------------------------
% \balance

\bibliographystyle{plain}
\bibliography{references}
%-------------------------------------------------------------------------------
%-------------------------------------------------------------------------------
% \newpage
%-------------------------------------------------------------------------------
%-------------------------------------------------------------------------------
\appendix
%-------------------------------------------------------------------------------
%-------------------------------------------------------------------------------
\section{Oblivious Transfer}
\label{sec:ot}
%-------------------------------------------------------------------------------
%-------------------------------------------------------------------------------
Oblivious Transfer (OT), first introduced by Rabin~\cite{rabin1981ot}, is a protocol that enables the exchange of secret messages between two parties. In a $b$ bit $n$-choose-$1$ OT, denoted by $\OT{N}{b}$, one party (sender) inputs a set of $n$ $b$ bit values $\{x_0, x_1, \cdots, x_{n-1}\}$, and the other party (receiver) inputs a choice $c \in [0, n)$ corresponding to an element in the set. The receiver learns only one of the sender's inputs, $x_c$ corresponding to its selection $c$, oblivious to the sender who does not learn anything. Correlated OT (COT) and Random OT (ROT) are two simpler types of OT that are powerful cryptographic primitives. In $\COT{2}{b}$, the sender inputs a private $b$ bit correlation $\delta$, a $b$ bit value $m$, and doesn't learn anything, while the receiver only learns $b$ bit $m+c\cdot\delta$. In $\ROT{2}{b}$, the sender does not input anything and learns $2$ random $b$ bit values $\{r_0, r_1\}$ while the receiver learns only one of them, $r_c$ for its input choice $c$. 

Notice, a $\ROT{2}{b}$ can generate a $\OT{2}{b}$, the server uses $r_0$ and $r_1$ as one-time-pads to mask $m_0$ and $m_1$, sends them to the receiver and the receiver uses $r_c$ to unmask and learn $m_c$. OT variants that assign \emph{random inputs} are core to generating correlated randomness for secure computation. In a \emph{random-choice} COT, the protocol automatically sets a random choice $c \in \mathbb{Z}_2$ and random message $m$, returns $m$ to the sender and the corresponding random string $m_c = r + c\cdot\delta$ and $c$ to the receiver. A \emph{random-choice} $\ROT{2}{b}$ can be generated from a random-choice $\COT{2}{b}$ by using a cryptographic hash function; the sender hashes $m$, $m + \delta$ to obtain $r_0$, $r_1$ respectively and the receiver hashes its COT output to get $r_c$.

Initial constructions of OT utilized public-key cryptography which incurred large overheads in terms of both computation and communication. OT extension protocols, first conceptualized by Beaver~\cite{beaver1996correlated} and later realized by Ishai et al.~\cite{ishai2003extending} (IKNP), reduce this overhead by generating a $O(n)$ OTs using lightweight symmetric key cryptographic operations from $O(n/k)$ public-key OTs (\emph{base OT}) where $k$ is a fixed constant parameterized by the OT extension protocol. Recently, Boyle et al.~\cite{boyle19SilentOT} devised a new OT extension protocol, namely \emph{silent OT extension}, which significantly reduces the number of base OTs required to $O(\log n)$ at the expense of more local computation.
%-------------------------------------------------------------------------------
\subsection{Silent OT Extension}
\label{sec:silentot}
%-------------------------------------------------------------------------------
The core components of Silent OT Extension are Punctured Pseudorandom Function (PPRF) and Encoding with Learning Parity with Noise (LPN). A Pseudorandom Function (PRF) is a deterministic random function that maps random values from a small domain to pseudorandom numbers in a larger domain. A \emph{Punctured} PRF (PPRF)~\cite{kiayias2013delegatable, boneh2013constrained} can be evaluated on all points in the original PRF's domain except $\alpha$, where it is punctured and returns $0$ (zero).

\begin{figure}[t]
    \centering
    \includegraphics[width=\columnwidth]{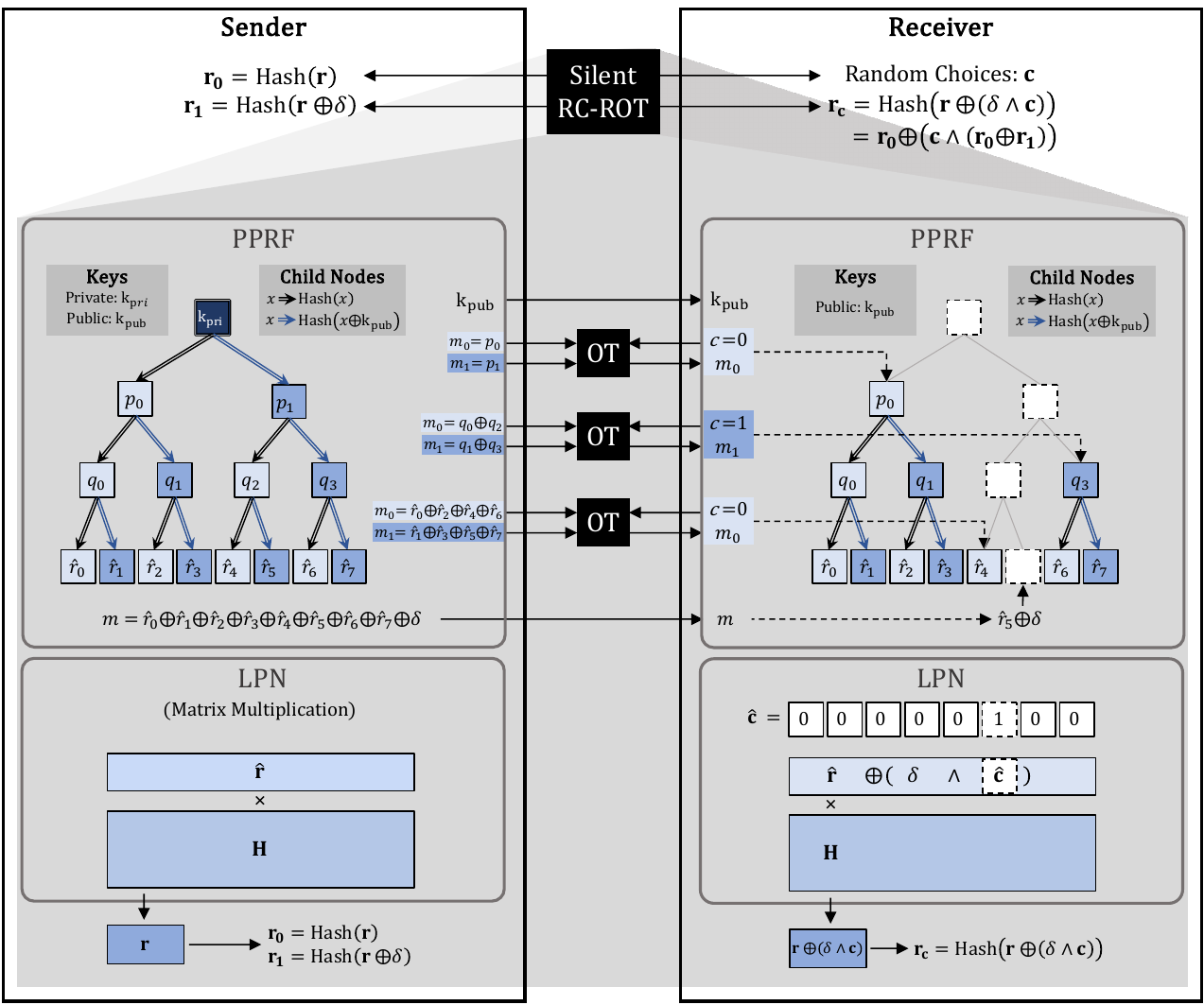}
    \caption{High-level illustration of Silent ROT protocol.}
    \label{fig:silentot}
\end{figure}

The LPN (dual) assumption~\cite{blum1993cryptographic} states that $(\mtrx{H}, \mtrx{H}\cdot\vctr{r})\stackrel{c}{\approx}(\mtrx{H},\vctr{b})$: the product of a known code matrix $\mtrx{H} = \mtrx{G}^{-1} \in \mathbb{Z}^{n\times 2n}_2$ and a very sparse random vector $r\in \mathbb{Z}^{2n}_2$ is computationally indistinguishable from a uniformly random vector $b\in \mathbb{Z}^n_2$. In Silent OT, LPN is used to compress $\hat{\vctr{r}}$, $\hat{\vctr{c}}$ and generate the $n$ COTs corresponding to the choice vector $\vctr{c}$. This involves computing a matrix-vector product with a very large matrix -- $n$ is in the order of $10^8$ for secure inference. To convert the COTs to ROTs, the sender computes two hashes of its output, while the receiver computes one hash. 

Overall, while silent OT offers significant gains in communication, it involves more intensive computation compared to IKNP-style protocols. The primary bottleneck in the computation arises from LPN encoding which involves a complexity of $O(n^2)$ that is quadratic in the size of the required number of OTs, $n$.

%-------------------------------------------------------------------------------
\subsection{Secure AND with Bit Triples}
\label{sec:secand}
%-------------------------------------------------------------------------------

In secure multiparty computation, evaluating the logical AND ($\land$) operation securely is a fundamental primitive. One efficient method to achieve this is by using Beaver's bit triples. A Beaver's bit triple consists of shared random bits ${\langle a \rangle_p^2, \langle b \rangle_p^2, \langle c \rangle_p^2}$ for each party $\mathcal{P}_p$ ($p \in {0:\text{Server}, 1:\text{Client}}$), such that $c = a \land b$ holds in the shared domain.

A bit triple can be constructed using two Random-Choice ROTs (RC-ROT), where each is a $\ROT{2}{1}$. The construction process is as follows:

\begin{algorithm}[t]
\setstretch{1.6} % Set line spacing
\caption{$\prtcl{AND}$ And with Beaver's Bit Triples}
\label{algo:and}
\DontPrintSemicolon
\KwInput{Input secret shares $\sshr{x}{p}{2}$ and $\sshr{y}{p}{2}$}
\KwOutput{Output secret share $\sshr{z}{p}{2}$, s.t. $z=x\land y$}
\tcc{\hfill \underline{\textnormal{Triple Retrieval}}}
$\Big\{\sshr{a}{p}{2}, \sshr{b}{p}{2}, \sshr{c}{p}{2}\Big\} = \mathsf{TripleGen.get}(1)$ \\
\tcc{\hfill \underline{\textnormal{Correction bit Computation}}}
$\sshr{e}{p}{2} = \sshr{a}{p}{2} \oplus \sshr{x}{p}{2}$ \\
$\sshr{f}{p}{2} = \sshr{b}{p}{2} \oplus \sshr{y}{p}{2}$ \\
\tcc{\hfill \underline{\textnormal{Correction bit Reveal}}}
$\mathsf{Send}(\sshr{e}{p}{2})$ ; $\mathsf{Send}(\sshr{f}{p}{2})$ \\
$\sshr{e}{p'}{2} = \mathsf{Receive}()$ ; $\sshr{f}{p'}{2} = \mathsf{Receive}()$\\
$ e = \sshr{e}{p}{2} \oplus \sshr{e}{p'}{2}$ ; $ f = \sshr{f}{p}{2} \oplus \sshr{f}{p'}{2}$\\
$\sshr{z}{p}{2} = (p' \land e \land f) \oplus (e \land \sshr{b}{p}{2}) \oplus (f \land \sshr{a}{p}{2}) \oplus \sshr{c}{p}{2}$
\end{algorithm}

\begin{enumerate} 
\item \textbf{First RC-ROT Instance}: 
\begin{itemize} 
    \item The server acts as the receiver, obtaining a random choice bit $d \in \mathbb{Z}_2$ and the corresponding random message $r_d$. 
    \item The client acts as the sender, obtaining random messages $r_0$ and $r_1$. 
    \item The client sets its share of $b$ as $\langle b \rangle_1^2 = r_0 \oplus r_1$. 
    \item The server sets its share of $a$ as $\langle a \rangle_0^2 = d$. 
    \item The relationship $r_d = r_0 \oplus (\langle a \rangle_0^2 \land \langle b \rangle_1^2)$ holds. 
\end{itemize} 

\item \textbf{Second RC-ROT Instance}: 
\begin{itemize} 
    \item The client acts as the receiver, obtaining a random choice bit $e \in \mathbb{Z}_2$ and the corresponding random message $s_e$. 
    \item The server acts as the sender, obtaining random messages $s_0$ and $s_1$. 
    \item The server sets its share of $b$ as $\langle b \rangle_0^2 = s_0 \oplus s_1$. 
    \item The client sets its share of $a$ as $\langle a \rangle_1^2 = e$. 
    \item The relationship $s_e = s_0 \oplus (\langle a \rangle_1^2 \land \langle b \rangle_0^2)$ holds. 
\end{itemize} 
\item \textbf{Computing Shares of $c$}: 
\begin{align} 
\langle c \rangle_0^2 &= \langle a \rangle_0^2 \land \langle b \rangle_0^2 \oplus r_d \oplus s_0 \\
\langle c \rangle_1^2 &= \langle a \rangle_1^2 \land \langle b \rangle_1^2 \oplus s_e \oplus r_0 
\end{align} 
\end{enumerate}

These computations ensure that when the parties combine their shares, they obtain $c = a \land b$, while keeping their individual inputs private.

Algorithm~\ref{algo:and} presents the protocol for the secure AND functionality. This algorithm is a straightforward implementation of the GMW protocol~\cite{goldreich1987play} utilizing preprocessed Beaver's bit triples. Both parties first retrieve a preprocessed bit triple to start the computation. They then compute local correction bits by XORing ($\oplus$) their input shares with the triple shares. These correction bits are exchanged and combined to reconstruct $e$ and $f$. Finally, each party computes its output share $\langle z \rangle_p^2$ using the reconstructed correction bits, the triple shares, and the bit-complement of party index $p'$, i.e. $p' = 1 - p$, ensuring that the underlying secret value in the shares is the logical AND of the inputs. 

This protocol allows the parties to compute the AND of their secret-shared bits without revealing their private inputs, forming a building block for more complex secure computations.

%-------------------------------------------------------------------------------
%-------------------------------------------------------------------------------
\section{Additional NN Operations}
%-------------------------------------------------------------------------------
%-------------------------------------------------------------------------------
\subsection{Max Pooling}
\label{sec:maxpool}
%-------------------------------------------------------------------------------
Pooling is a fundamental operation in CNNs aimed at downsampling input feature maps to highlight dominant features. Max Pooling achieves this by extracting the maximum value from each specified segment of the input array. In \fw{SecONNds}, Max Pooling is implemented securely through a protocol called $\prtcl{MaxPool}$, which is outlined in Algorithm~\ref{algo:maxpool}.

The core functionality of the $\prtcl{MaxPool}$ protocol involves taking secret shares of an input layer and securely determining the maximum values for designated regions or windows. The protocol initializes the presumed maximum with the first element of each window and securely iterates through the remaining elements to find the actual maximum. Each comparison is performed with one call to $\prtcl{ReLU}$, totaling $w$ calls to compute one output element where $w$ is the flattened window size for pooling. 

\begin{algorithm}[t]
\setstretch{1.6} % Set line spacing
\caption{$\prtcl{MaxPool}$ Max Pooling}
\label{algo:maxpool}
\DontPrintSemicolon
\SetAlgoVlined
    \KwInput
    {Flattened window size $w$\newline
    Input secret share array $\sshr{\vctr{i}}{p}{2^b}$ of size $w$
    }
    \KwOutput{output secret share $\sshr{o}{p}{2^b}$}
    
    $\sshr{o}{p}{2^b} = \sshr{\vctr{i}}{p}{2^b}[0]$\\
    
    \For{$k = 1$ \KwTo $w-1$}
    {
        $\sshr{o}{p}{2^b} = \prtcl{ReLU}\Big(\sshr{o}{p}{2^b} - \sshr{\vctr{i}}{p}{2^b}[k]\Big) + \sshr{\vctr{i}}{p}{2^b}[k]$
    }
\end{algorithm}
%-------------------------------------------------------------------------------
\subsection{Average Pooling}
\label{sec:avgpool}
%-------------------------------------------------------------------------------
Average pooling is an operation commonly used in neural networks to reduce the spatial dimensions of feature maps while retaining important information. In a secure inference setting, we need to perform average pooling on secret-shared data without revealing the underlying values. Algorithm~\ref{algo:avgpool} presents the protocol for average pooling adapted from CrypTFlow2~\cite{rathee20cryptflow2}, which is employed in our system. 

The protocol operates on secret shares of the input and produces secret shares of the output. The input matrix $\langle \mtrx{I} \rangle_p^{2^b}$ contains secret shares of the input values, where $b$ is the bit width. The protocol initializes the output vector by assigning the first element from each window. It then iteratively accumulates the remaining elements in each window by performing secure addition on the secret shares. Finally, the protocol performs a division by the window size $w$ using the $\mathsf{DIV}$ protocol from CrypTFlow2, which securely computes the division of secret-shared values by a public constant.

The division protocol $\mathsf{DIV}$ takes as input secret shares of the dividend and a public divisor and returns secret shares of the quotient. This operation is performed securely without revealing the intermediate sums or the final averaged values. By employing this protocol, we securely compute the average pooling operation on secret-shared data, which is essential for privacy-preserving neural network inference.

\begin{algorithm}[t]
\setstretch{1.6} % Set line spacing
\caption{$\prtcl{AvgPool}$ Average Pooling}
\label{algo:avgpool}
\DontPrintSemicolon
\KwInput{Output size $n$ ; Flattened window size $w$\newline
Input secret share array $\sshr{\mtrx{I}}{p}{2^b}$ of size $n \times w$
}
\KwOutput{Flattened output secret share $\sshr{\vctr{o}}{p}{2^b}$}

\For{$j = 0$ \KwTo $n-1$}
{
$\sshr{\vctr{o}}{p}{2^b}[j] = \sshr{\mtrx{I}}{p}{2^b}[j, 0]$
}
\For{$k = 0$ \KwTo $w-2$}
{
\For{$j = 0$ \KwTo $n-1$}
{
$\sshr{\vctr{o}}{p}{2^b}[j] = \sshr{\vctr{o}}{p}{2^b}[j] + \sshr{\mtrx{I}}{p}{2^b}[j, k]$
}
}
$\sshr{\vctr{o}}{p}{2^b}[j] = \prtcl{DIV}(\sshr{\vctr{o}}{p}{2^b}, w)$
\end{algorithm}
%-------------------------------------------------------------------------------
%-------------------------------------------------------------------------------
\section{RLWE HE}
\label{sec:appendix_he}
%-------------------------------------------------------------------------------
%-------------------------------------------------------------------------------
Homomorphic encryption (HE) enables computations on encrypted data without requiring decryption keys or access to the original plaintext. This approach allows a client to encrypt data, send it to a server for computation, and then decrypt the results. In HE systems, encryption and decryption are managed by the client, while the majority of the computational work is offloaded to the server. This division of labor reduces communication overhead, with data transfer involving only input and output sizes. Fully Homomorphic Encryption (FHE) schemes, introduced by Gentry \cite{gentry09fhe}, support arbitrary encrypted computations through a bootstrapping process but come with significant computational costs. Leveled HE schemes improve computational efficiency by allowing a limited number of operations without bootstrapping, although they increase ciphertext size. The capacity of these schemes is determined by their multiplicative depth, which indicates the maximum number of sequential multiplications.

The construction of most popular HE schemes, such as BGV \cite{bgv11}, BFV \cite{brakerski11bfv, fan12bfv}, and CKKS \cite{cheon17ckks}, relies on \emph{Ring Learning With Errors} (RLWE)~\cite{lyubashevsky13rlwe}. RLWE-based HE schemes operate on polynomial rings, denoted as $\mathcal{R}^{N}_{\,\,\,Q}$, where $N$ represents the \emph{polynomial modulus degree} and $Q$ represents the \emph{coefficient modulus}. In schemes such as BFV and BGV, plaintexts are elements of $\mathcal{R}^{N}_{\,\,\,P}$, with $P$ ($<Q$) serving as the plaintext modulus. In CKKS, plaintexts are elements in $\mathcal{R}^{N}_{\,\,\,Q}$, without a plaintext modulus, facilitating approximate arithmetic.

Encryption in RLWE-based schemes involves encoding a secret vector $\vctr{m} \in \mathbb{Z}_P^N$ into a polynomial $\pt{m}$. The encryption yields a ciphertext $\ct{m} = (\poly{a}, \poly{b})$, where $\poly{a}$ is a random polynomial in the polynomial ring $\mathcal{R}^{N}_{\,\,\,Q}$, and $\poly{b}$ is a polynomial in $\mathcal{R}^{N}_{\,\,\,Q}$ s.t.:

\[
\poly{b} = \left(\poly{a}\cdot\poly{sk} + \delta\cdot\pt{m} + \epsilon\cdot\poly{e}\right)
\]

Here, $\poly{sk} \in \mathcal{R}^{N}_{\,\,\,\{-1, 0, 1\}}$ represents the secret key, which is a polynomial with uniformly random ternary coefficients, $\delta, \epsilon \in \mathbb{Z}_{Q}$ are scheme-defined constants used to scale the plaintext and control noise tolerance, respectively, and $\poly{e} \in \mathcal{R}^{N}_{\,\,\,Q}$ is a small error polynomial drawn from a zero-mean discrete Gaussian distribution with standard deviation $\sigma$. This small error $\poly{e}$ is crucial for maintaining the hardness of the RLWE problem, thus ensuring the security of the encryption.

In HE, operations are performed on the ciphertexts by manipulating the underlying polynomials. Addition, multiplication, and rotation are the key operations supported by RLWE schemes. Addition and rotation lead to additive error growth, whereas multiplication results in multiplicative error growth, making parameter selection critical to control the error expansion.

HE multiplication corresponds to computing a convolution of the polynomial coefficients. Specifically, multiplying two polynomials $\poly{a}$ and $\poly{b}$ yields a result $\poly{c} = \poly{a} \times \poly{b}$, where each coefficient $c_i$ is determined through the convolution of the coefficients of $\poly{a}$ and $\poly{b}$:

\[
c_i = \sum_{j=0}^{i} a_j \cdot b_{i-j} \quad \text{(mod $Q$)}
\]

This direct convolution has a quadratic complexity of $O(N^2)$, which can be computationally prohibitive for polynomials of high degree.
%-------------------------------------------------------------------------------
\subsection{Number Theoretic Transform}
\label{sec:ntt}
%-------------------------------------------------------------------------------
To alleviate the aforementioned quadratic computational complexity, the \emph{Number Theoretic Transform} (NTT) is used, which is analogous to the Fast Fourier Transform (FFT) but tailored for modular arithmetic, suitable for cryptographic applications. The NTT allows polynomials to be transformed into a point-value representation, where convolution is transformed into pointwise multiplication. This reduces the complexity of polynomial multiplication from $O(N^2)$ to $O(N \log N)$. The NTT uses a "primitive root of unity" under modulo $Q$, unlike the FFT, which employs complex roots of unity.

NTT-based multiplication proceeds in three main steps. First, both polynomials $\poly{a}$ and $\poly{b}$ are transformed using the NTT, resulting in $\text{NTT}(\poly{a})$ and $\text{NTT}(\poly{b})$. The transformation is applied to convert the polynomial from its coefficient representation to a point-value representation in $\mathcal{R}^{N}_{\,\,\,Q}$. Second, pointwise multiplication is performed -- one scalar multiplication per coefficient:

\[
\text{NTT}(\poly{c})_i = \text{NTT}(\poly{a})_i \cdot \text{NTT}(\poly{b})_i \quad \text{(mod $Q$)}
\]

\noindent
Finally, an inverse NTT is applied to obtain the product $\poly{c}$ back in the coefficient domain, $\poly{c} = \text{NTT}^{-1}(\text{NTT}(\poly{c}))$.

The use of the NTT not only reduces computational costs but also ensures that operations remain consistent with the modular arithmetic required for cryptographic security. The advantage of the NTT in HE is that it optimizes the core operation of polynomial multiplication, transforming the otherwise convolution-heavy multiplication into a sequence of efficient element-wise multiplications. Given that polynomial multiplication is central to encrypted computation, this optimization is essential for making homomorphic encryption schemes practical and efficient. The use of modular arithmetic throughout ensures that all operations are secure and stay within the defined finite field, which is critical for maintaining the integrity and security of HE schemes.
%-------------------------------------------------------------------------------
%-------------------------------------------------------------------------------
\subsection{ResNet50 Evaluations}
\label{sec:resnet}
\begin{figure}[t]
    \centering
    \includegraphics[width=\columnwidth]{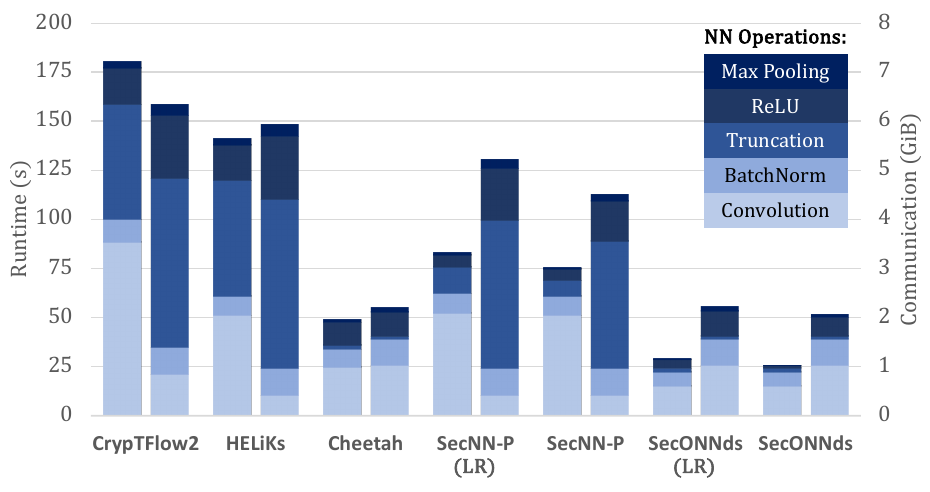}
    \caption{End-to-end (E2E) runtime (left bar) and communication (right bar) performance of each framework for Neural Network operations in ResNet50.}
    \label{fig:resnet}
\end{figure}

%-------------------------------------------------------------------------------
%-------------------------------------------------------------------------------
In Figure~\ref{fig:resnet} we show the results of our evaluations on the ResNet50 model with $32$ Threads. \fw{SecONNds} significantly outperforms existing frameworks in both runtime and communication efficiency. Specifically, \fw{SecONNds} achieves a total runtime of $25$ seconds, demonstrating a $1.9\times$ speedup over \fw{Cheetah}'s $49$ seconds and a substantial $7\times$ improvement compared to \fw{CrypTFlow2}'s $180$ seconds. In terms of communication volume, \fw{SecONNds} reduces data transfer to $2061$ MiB, marking an $8.5\%$ decrease from \fw{Cheetah}'s $2248$ MiB. 

The \fw{SecONNds\text{-}P} variants maintain competitive performance, balancing runtime and communication efficiency across different configurations. These results underscore \fw{SecONNds}' effectiveness in optimizing secure inference for complex neural network architectures like ResNet50.

%-------------------------------------------------------------------------------
%-------------------------------------------------------------------------------

%%%%%%%%%%%%%%%%%%%%%%%%%%%%%%%%%%%%%%%%%%%%%%%%%%%%%%%%%%%%%%%%%%%%%%%%%%%%%%%%
\end{document}